\documentclass[twoside,twocolumn,9pt]{article}
\usepackage{extsizes}
\usepackage{natbib} 

\usepackage[version=3]{mhchem}
\usepackage[left=1.5cm, right=1.5cm, top=1.785cm, bottom=2.0cm]{geometry}
\usepackage{balance}
\usepackage{sectsty}
\usepackage{graphicx} 
\usepackage[format=plain,justification=justified,singlelinecheck=false,font={stretch=1.125,small,sf},labelfont=bf,labelsep=space]{caption}
\usepackage{float}
\usepackage{fnpos}
\usepackage[english]{babel}
\addto{\captionsenglish}{%
  
}
\usepackage{array}
\usepackage{charter}
\usepackage[T1]{fontenc}
\usepackage[usenames,dvipsnames]{xcolor}
\usepackage{setspace}
\usepackage[compact]{titlesec}
\usepackage{hyperref}

\usepackage{bm}
\usepackage{amssymb}

\newcommand{\Ab}{\mathbf{A}}
\newcommand{\Bb}{\mathbf{B}}
\newcommand{\Cb}{\mathbf{C}}
\newcommand{\Db}{\mathbf{D}}
\newcommand{\Gb}{\bm{\Gamma}}

\usepackage{authblk}

\usepackage{epstopdf}

\definecolor{cream}{RGB}{222,217,201}

\title{Exponential speed-up in VQE molecular energy ranking with Sridhara-compressed Hamiltonians
} 

\author[1]{Dennis Lima}
\author[1]{Saif Al-Kuwari} 
\affil[1]{Qatar Center for Quantum Computing, College of Science and Engineering, Hamad Bin Khalifa University}

\begin{document}
 \maketitle
 
\begin{abstract}
Polycyclic aromatic hydrocarbons (PAHs) are residual and intermediary molecules in the Chemical Vapor Deposition (CVD) to produce graphene from methane. Ranking a combinatorial space of variants of PAHs by energy allows the CVD to be optimized, while simulations of PAHs are strong candidates for quantum advantage in quantum computers.
We extend on Sridhara's root formula to perform block diagonalization (SBD) of six PAHs using Hartree-Fock Hamiltonians with STO-3G basis set and $(2,2)$, $(4,4)$, $(6,6)$ settings of active orbitals and active electrons. 
We show that the proposed SBD algorithm followed by Variational Quantum Eigensolver (VQE) allows ranking molecules by ground state energy with $77.8\%$ of success in comparison with the uncompressed VQE, while speeding up the VQE simulation in $164.16\%$ (median) keeping its average error of active space reduction down to $0.09\%$. We conclude that the flexibilization of constraints of the SBD algorithm makes it a fast and reliable estimator for active space reduction in molecular simulation.
\end{abstract}

\footnotetext{\textit{$^{a}$~Qatar Center for Quantum Computing, College of Science and Engineering, Hamad Bin Khalifa University, Doha, Qatar.}}
\footnotetext{\textit{$^{*}$~E-mail: deaq54989@hbku.edu.qa}}

\section{Introduction} \label{sec:intro}

The Noise Intermediate-Scale Quantum (NISQ) era is marked by advances in pre-processing, error mitigation and error correction in quantum optimization algorithms in benchmarking and scalability studies. As the quantum advantage era approaches, Hamiltonian preparation and new quantum eigensolvers become major opportunities for cutting-edge discoveries in quantum chemistry\cite{gemeinhardt2023quantum} for carbon mitigation and economic growth\cite{shukla2022climate}.  

In the methane cycle, methane emissions account for about $30\%$ of global warming, with enteric fermentation and leakage from oil production being the two main sources\cite{shukla2022climate}. While research in precision fermentation and alternative protein seek to improve food security and viabilize the transition to a vegan lifestyle as the most efficient way to reduce the leading source of methane \cite{malila2024current}, mitigation of the second main source relies on research of methods to convert natural gas into greener valuable materials, like graphene compounds \cite{CAO2023100522} and hydrogen gas \cite{patlolla2023review}.

Chemical vapor deposition (CVD) is the most scalable process to synthesize graphene sheets using methane as precursor \cite{naghdi2018catalytic, saeed2020chemical}. The high sensitivity of two-dimensional carbon lattices during their growth process results in the residual formation of coke, carbon nanotubes and polycyclic aromatic hydrocarbons (PAHs)  \cite{norinaga2007detailed, lu2015molecular, cao2023research}. A better understanding of their energy profiles that leads to their ranking from most to least abundant would strongly contribute to mitigate catalyst poisoning and improve conversion efficiency \cite{tong2022decarbonizing}. 
    \begin{figure}[ht]
        \centering
         \includegraphics[clip, trim= 1.9cm 0 0 0,width=1\linewidth]{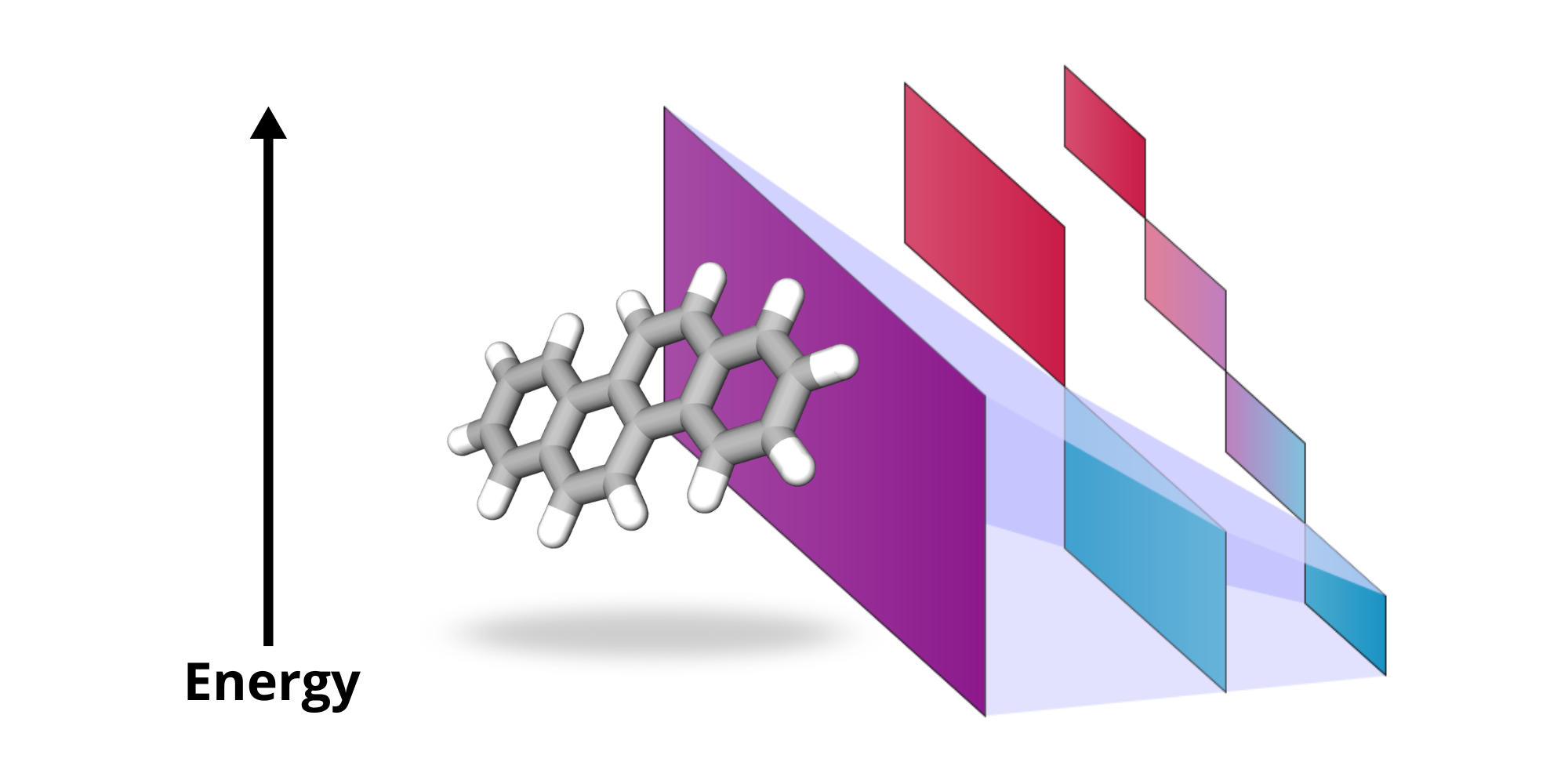}
        \caption{The block diagonalization procedure rotates the molecular Hamiltonian into a high-energy and a low-energy block at each step of the iteration. }
        \label{fig1}
    \end{figure}

In this realm, tetracyclic aromatic hydrocarbons (TAHs) are valuable candidates to quantum advantage due to the occurrence of strong electronic correlations \cite{pelzer2011strong} that are not efficiently simulated in classical computers due to Hamiltonian size\cite{tilly2022variational}. 

The number of quantum operations and the number of qubits are the most relevant scale properties to adapt Hamiltonians to quantum eigensolvers. Standard techniques for reducing circuit depth include Trotterization\cite{PhysRevA.106.042401}, truncation after factorization\cite{Zhong2021Quantum}, while the limiting number of qubits is overcome by compression techniques such as tapering\cite{bravyi2017tapering}, lossy quantum annealing compression\cite{yoon2022lossy}, space reduction\cite{arnoldi1951principle,bavely1979algorithm} and block diagonalization\cite{DAVIDSON197587, ghani2020novel}.

\begin{table*}[ht] \small
\caption{\ Most used algorithms to solve matrix polynomials and the block eigenvalue problem. N/A stands for "not applicable"}
\label{table1}
\begin{tabular}{llp{1.6cm}p{4.6cm}p{0.95cm}p{1.5cm}p{1.45cm}}
\hline
Algorithm                    &Problem                       & Best-Case Convergence& Limitations                                                                                      & Proposal (Scalar)                                    & Generalization (Matrix)& Latest improvement\\ \hline
Sridhara&Solvent of matrix polynomial&             Quadratic   & Requires blocks to obey some commutativity constraints. Leading solvent is the most stable.      & c. 900 CE \cite{sridhara1959}                        & 2001\cite{higham2001solving}                                                              & 2025 (This paper)                                                                      \\
 Newton& Solvent of matrix polynomial  & Quadratic& Sensitive to spectral separation.& 1669\cite{Guicciardini_2016,tjalling1995}& 1981\cite{davis1981numerical,  davis1983algorithm}&2023\cite{higham2001solving, macias2023two}\\
 Bernoulli& Solvent of matrix polynomial  & Linear      & Slow, difficult dominant solvent verification.                                                   & 1728\cite{Bernoulli1728, dutka1995early}& 1971\cite{dennis1971matrix}&2025\cite{meyer2025solving, gohberg2009matrix,gohberg1982matrix}\\
Traub&Solvent of matrix polynomial  &             Linear      &                                                                                                  Slow, difficult dominant solvent verification.                                                   &                                                      1966\cite{traub1966class}& 1971\cite{dennis1971matrix}& N/A\\
                             Davidson&Block eigenvalue              & Superlinear& Better for semi-diagonal matrices, failing for diagonal matrices. Matrix must be real symmetric. &                                                      1975\cite{DAVIDSON197587,  morgan1986generalizations} & 1978\cite{ liu1978simultaneous, MURRAY1992382} &                                                                                        2016\cite{Parrish2016, leininger2001systematic}                                                     \\
 Bavely-Stewart& Block eigenvalue & N/A& May fail if spectrum is degenerate.& None& 1979\cite{bavely1979algorithm}                                                            &N/A\\
                             Jacobi&Block eigenvalue&             Quadratic&                                                                                                  Designed for Hermitian matrices only.&                                                      1993\cite{veselie1993jacobi}& 2014\cite{HARI20141}&                                                                                        2024\cite{BEGOVICKOVAC2024421,saad2023revisiting}\\ \hline
\end{tabular}
\end{table*}

Block diagonalization is (Fig. \ref{fig1}) the process of finding the complete set of block eigenvalues of a matrix $H$ , i.e. smaller square matrices $\Gb_i$ that preserve the eigenvalues of $H$  (Eq. \ref{eq1}), where $V$ is the block-eigenvector matrix. This method offers similar advantages to diagonalization for small block size, in addition to facilitating faster matrix power expansions \cite{bavely1979algorithm}.
\begin{align} \label{eq1}
    VH  = H_\text{BD}V.
\end{align}
The first general study of block eigenvalues was developed by Dennis, Traub and Weber in the theory of matrix polynomials\cite{dennis1976algebraic, dennis1978algorithms}. Right solvents of the matrix polynomial $P(X)$ are matrices $S$ that satisfy\cite{dennis1976algebraic} $P(S)=0$ (Eq. \ref{eq:poly}). If $P(X)$ represents a secular equation of a matrix $H$, then the solvents $S\in (\Gb_0, \Gb_1)$ are block eigenvalues of $H$.
\begin{align} \label{eq:poly}
P(X) = A_2X^2+ A_1X + A_0 = 0
\end{align}

Early attempts to solve matrix polynomials focused on generalizing scalar methods to block matrices, such as the Block Newton method\cite{davis1981numerical, davis1983algorithm} as a matrix version of the Newton method\cite{tjalling1995}, the Block Bernoulli method \cite{gohberg1982matrix, gohberg2009matrix} as a matrix version of the Bernoulli method \cite{Aitken_1927}, and the Block Sridhara \cite{higham2001solving} method (Eq. \ref{e:bsri}) as an adaptation to Sridhara's completion of squares \cite{sridhara1959}. Table \ref{table1} lists these and other methods that have been generalized or improved in recent years. We refer to the Block Sridhara method applied to a matrix secular equation as Sridhara Block Diagonalization (SBD) throughout this paper.

\begin{align} \label{e:bsri}
    X &= -\frac{1}{2}A_1+ \frac{1}{2}(A_1^2 - 4A_0)^{1/2} 
\end{align}

The solvent Eq. \ref{e:bsri} with $A_2=I$  is exact under several constraints \cite{higham2001solving}: $A_1$ and $A_0$ must commute (Eq. \ref{eq:cond1}), the square root must exist and, in some algorithms \cite{higham2001solving}, either the norm of $A_1$ or its inverse should be smaller than those of $A_2$ and $A_0$.

\begin{align}
    \label{eq:cond1} [A_1,A_0] &= 0,
\end{align}

Under these restrictions, the Block Sridhara method has limited applicability for exact results\cite{higham2001solving}, being mostly used in perturbative problems where either $A_2$ or $A_1$ has a small Frobenius norm \cite{higham2001solving}.

After thorough developments on the properties of matrix polynomials and block eigenvalues up to 2001 (Table \ref{table1}), optimizations of iteration methods became prevalent, especially the Newton method \cite{higham1986newton, higham2001solving}, Wielandt's deflation method \cite{PEREIRA20011177}, and the quantum natural gradient method \cite{stokes2020quantum}. The Lanczos (1950)\cite{lanczos1950iteration, zhang2001lanczos} and Arnoldi (1951) \cite{arnoldi1951principle} methods are subspace reduction methods that make use of tridiagonal and triangular transformations, but do not directly block-diagonalize matrices.

As an ubiquitous technique in modern dynamical problems, block diagonalization is used in estimation of antenna impedance matrix \cite{wu2021hybrid}, symmetry search in many-body Hamiltonians \cite{schmitz2020quantum}, and linear response equations in theoretical chemistry \cite{alessandro2023linear}.

Unlike most methods in Table \ref{table1}, the Block Sridhara computes solvents via an iteration to approximate a matrix square root, rather than directly iterating to obtain the block eigenvalue. As a consequence, quadratically-convergent matrix square root recursions like Newton's \cite{higham1986newton, higham1997stable} can make it outperform the majority of block diagonalization and solvent methods.

If the restrictions of Sridhara's method are flexibilized for arbitrary matrices, with parameters that allow a numerical error minimization, its quadratic convergence will make it a competitive general-purpose block diagonalization algorithm for ground state estimation. 

\begin{figure}[ht]
        \centering
        \includegraphics[width=1\linewidth]{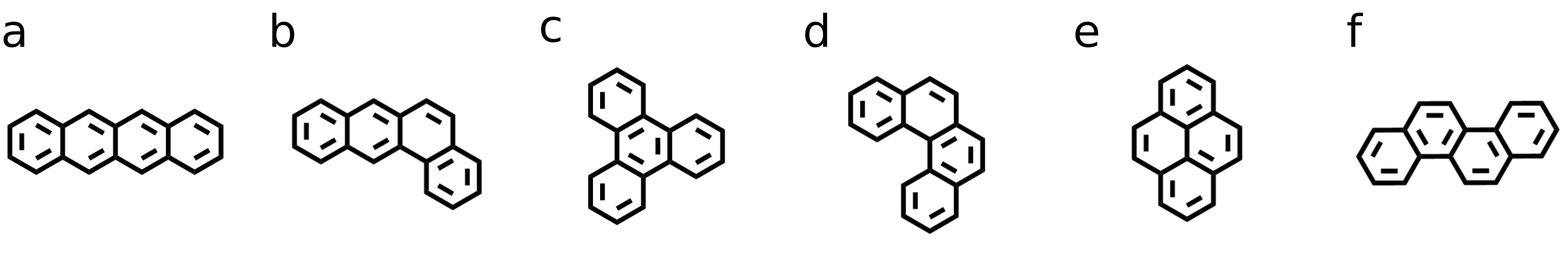}
        \caption{\textbf{Tetracene, benz(a)anthracene, triphenylene, benzo[c]phenanthracene, pyrene and chrysene, respectively, from a to f. TAHs are common residuals and precursors in graphene synthesis.}}
        \label{f:molecules}
    \end{figure}

In this paper, we investigate the stability and error propagation of a variant of the Sridhara-based block diagonalization followed by Arnoldi sparse eigensolver and Variational Quantum Eigensolver. As a benchmark of industrial relevance, we analyze the energy ranking of six tetracyclic aromatic hydrocarbons (Fig. \ref{f:molecules}) under different compression depths.

\section{Methods} \label{sec:methods}
    In this section, we introduce the techniques used to compute the Hamiltonians, compress them, and compute their ground states in \texttt{python} programming language (version 3.11.9). 
    
    \subsection{Hartree-Fock (HF) Hamiltonian preparation}
        We computed the HF Hamiltonians in STO-3G basis of tetracene (CID 7080  ), benz(a)anthracene (CID 5954), triphenylene (CID 9170), benzo[c]phenanthracene (CID 9136), pyrene (CID 31423) and chrysene (CID 9171), illustrated in Fig. \ref{f:molecules}), respectively, using the \texttt{pennylane} python package \cite{bergholm2018pennylane} to interface with \texttt{openfermion}\cite{mcclean2020openfermion} and \texttt{pyscf} \cite{sun2018pyscf}.
        The numbers of active electrons and active orbitals were set as $(2,2)$, $(4,4)$ and $(6,6)$ for each molecule, in order to analyze the influence of the scale of the active space on the performance of the compression algorithm.

        As each orbital in the Fermionic HF Hamiltonian is mapped to a two-qubit space after the Jordan-Wigner transformation, these active spaces correspond to 4-qubit, 8-qubit and 12-qubit Hamiltonians for each molecule, respectively. 
        
        The choice of STO-3G basis serves the algorithm for benchmarking purposes, rather than calculation of exact ground states, as this basis is of low accuracy.

    \subsection{Sridhara Block Diagonalization}
        The matrix form of Sridhara's method for the roots of a quadratic polynomial \cite{higham2001solving} also known as the \textit{method of completing the squares}, is described in Eqs. \ref{eq:sridhararoute} through \ref{eq:sridhararoute2}. 
        \begin{align} \label{eq:sridhararoute}
            X^2 + A_1X + A_0 &= 0
            \\ X^2 + A_1X + \left(\frac{1}{4} A_1^2 - \frac{1}{4} A_1^2 \right) + A_0 &= 0
            \\ \left(X + \frac{1}{2}A_1\right)^2 
            &= \frac{1}{4} A_1^2-A_0
            \\ X + \frac{1}{2}A_1 
            &= \left( \frac{1}{4} A_1^2-A_0 \right)^{1/2}
            \\ X 
            &= -\frac{1}{2}A_1 + \left(  \frac{1}{4}A_1^2-A_0 \right)^{1/2}
            \label{eq:sridhararoute2}
        \end{align}
     
        The signs of the eigenvalues are encoded within the square root but, if the radicand is positive-definite, then that root is also positive-definite (numerically). Therefore, the only accessible solutions to Eq. \ref{eq:sridhararoute2} are itself and Eq. \ref{eq:sridhararoute2} with a negative sign outside the root. These solvents form a complete set of block eigenvalues \cite{higham2001solving}.

        We proceed by defining the Hamiltonian $H$ as a block matrix and $\Gb I$ as its block eigenvalue matrix to be solved for $\Gb$.
        \begin{align} 
        \label{e:hamp2}
        H &= 
        \begin{pmatrix} 
           \Ab & \Bb
        \\ \Cb & \Db
        \end{pmatrix}, \quad \text{with} \quad \dim{ \Ab } = \dim{ \Db } = \frac{\dim{ H }}{2},
        \\ \label{e:hamg}
        \Gb I &=
        \begin{pmatrix} 
        \Gb & 0 \\ 0 & \Gb
        \end{pmatrix}.
        \end{align}

        Initially, we recall the scalar secular equation in terms of block determinants (Eqs. \ref{e:sec}) of matrices that are not necessarily commutative. 
        \begin{align} \label{e:sec}
        \mathrm{det} \, (H - \Gb I) &=0
        \\ \notag
        \leftrightarrow \quad 
        \mathrm{det}[\Ab-\Gb ]
        \mathrm{det}[\Db-\Gb
        - \Cb(\Ab-\Gb )^{-1}\Bb] &= 0.
        \end{align}

        Our first approximation begins with adopting $\mathrm{det_{block}}$ as a flexibilization of the block determinant $\mathrm{det}$ for block matrices in place of scalars  \cite{Silvester_2000}. This leads to Eq. \ref{e:firstapprox}, a non-commutative secular equation.
        \begin{align} \label{e:firstapprox}
        \mathrm{det_{block}} (H - \Gb I) &= 0
        \\  \notag
        \leftrightarrow \quad (\Ab-\Gb )(\Db-\Gb ) 
        - (\Ab-\Gb ) \Cb(\Ab-\Gb )^{-1}\Bb &= 0.
        \end{align}

Using the substitution $\Cb^\prime = (\Ab-\Gb ) \Cb(\Ab-\Gb )^{-1}$ and expanding Eq. \ref{e:firstapprox}, one obtains Eq. \ref{eq:expanded}.

\begin{align} \label{eq:expanded}
\Gb^2 - \Ab\Gb -\Gb\Db + \Ab\Db -\Cb^\prime \Bb =0
\end{align}

Assuming $[D, \Gb]= 0$, one obtains Eq. \ref{eq:newt}.

\begin{align} \label{eq:newt}
    \Gb^2 - (\Ab + \Db)\Gamma + \Ab\Db -\Cb^\prime \Bb  =0
\end{align}

Using the substitutions $X=\Gb$ ,  $A_1=-(\Ab + \Db)$ , and $A_0=\Ab\Db -\Cb^\prime \Bb $, in addition to the approximation $\Cb^\prime \approx \Ab\Cb\Ab^{-1}$, Eq. \ref{eq:newt}  becomes Eq. \ref{eq:poly} , which is solved into Eq. \ref{broot0} the same way as Eq. \ref{eq:sridhararoute2}. Here, $-(-1)^\alpha$ is a convenience to index the sign of the root, with $\alpha\in \{0,1\}$.
      
        \begin{align} \label{broot0}
            \Gb_\alpha    = \frac{1}{2}(\Ab+\Db) -(-1)^\alpha \left[\frac{1}{4}(\Ab+\Db)^2 - (\Ab\Db-\Ab\Cb\Ab^{-1}\Db)\right]^{1/2}
            .
        \end{align}

Although Eq. \ref{broot0} is an exact solvent for a matrix equation of the form of Eq. \ref{eq:sridhararoute}, the flexibilization from Eq. \ref{eq:newt} through Eq. \ref{broot0rec} make it bear an inherent error due to a hidden variable.

Eq. \ref{eq:cond1} is satisfied or if
$\Gb$, $\Ab$, $\Bb$, $\Cb$ and $\Db$ are simultaneously diagonalizable, which makes $\Ab\Cb\Ab^{-1}=\Cb$, the similarity transformation with $\Ab$ reduces the error that arises from loss of commutativity due to the square root being a numerical approximation. 

Since $H$ is rescaled into a positive semi-definite matrix, then $\Gb_1$ is the leading solvent and contains the global top state and the global above-median state as its ground state, while $\Gb_0$ contains the global ground state. In case $\Ab$ is not invertible, equation $\det^\prime(H) = \Ab \Db - \Cb\Bb $ is used as an approximation instead of $\det_\mathrm{block}(H)$.

The recursive version of Eq. \ref{broot0} is Eq. \ref{broot0rec}, with the index $b\oplus\alpha$ as a bit string composition of a bit string $b$ and a new bit $\alpha$. For example, after two steps of block diagonalization, $H$ will be composed of four blocks ($\Gb_{00},$ $\Gb_{01},$ $ \Gb_{10},$ $ \Gb_{11}$), where the left-hand index is $b$ and the right-hand index is $\alpha$.

\begin{align}
    \label{broot0rec}
    \Gb_{b \oplus \alpha}    = \frac{1}{2}\mathrm{Tr_{block}}(\Gb_b) -(-1)^\alpha \left[\frac{1}{4}(\mathrm{Tr_{block}}(\Gb_b))^2 - \mathrm{det_{block}}(\Gb_b)\right]^{1/2}
\end{align}

    In order to privilege scalability, we avoid eigensolving the square root in Eq. \ref{broot0rec} and opt for truncating the Newton-Schulz expansion \cite{STOTSKY2020883} at the sixth term (Eqs. \ref{e:nsch}, \ref{e:nsch1}, \ref{e:nsch2} with $r=5$, and $M$ as an arbitrary matrix), since convergence arises at this point for most matrices. The coefficient of the identity matrix $I$ in $K_0$ in Eq. \ref{e:nsch2} was chosen heuristically as the fourth root of the trace for better convergence.

        \begin{align} \label{e:nsch}
            M^{1/2}    &= \lim_{r\to\infty} K_{r+1},
            \\ \label{e:nsch1}
            \text{with} \quad  K_{r+1} &= \frac{1}{2} (K_r + MK_r^{-1}),
            \\ \label{e:nsch2}
            \text{and} \quad K_{0} &= (\mathrm{Tr} M)^{1/4} I.
        \end{align}
        
        The deviation from Hermiticity that arises from such approximations is mitigated by using the Hermitization in Eq. \ref{e:herm}. and a set of correction parameters $N$ and $T$ at each step to force the eigenvalues to remain between $0$ and $1$ for a more stable eigensolving (Eq. \ref{e:adjust}).
        \begin{align} \label{e:herm}
            \Gb_{b}^{\prime} &= \Gb_b \Gb_{b}^{\dagger}
            \\ \label{e:adjust}
            \Gb_{b}^{\prime\prime} &= \frac{1}{N}\Gb_{b}^{\prime} + TI 
        \end{align}

At the last step, after eigensolving the last block eigenvalue, the resulting eigenvalue $\epsilon$ undergoes reverse operations (Eq. \ref{e:rev2}) to recover the original eigenvalue, $\gamma$ in Eq. \ref{e:rev}. The parameters $N_0$, $T_0$, $N$ and $T$  are chosen based on a previous guess of the magnitude and sign of the eigenvalues so that the eigensolvers only solve positive numbers between $0$ and $1$ to ensure optimal stability. 
        \begin{align} \label{e:rev}
            \gamma(m) &= (\epsilon_{m}-T_0)N_0,
            \\ \label{e:rev2} \mathrm{with} \quad \epsilon_{k} &=  \left| \sqrt{(\epsilon_{k-1}-T)N} \right|,
            \\ \label{e:rev3} \text{and with} \quad\epsilon_0 &=\left| \mathrm{Eigensolve}(\Gb_b) \right|.
        \end{align}
               
        Although $N_0$, $N$,  $T_0$ and $T$ seem to be arbitrary, most molecular Hamiltonian models have a ground state or top state in a predictive domain, and this domain helps select appropriate values of   $T$ and $N$. For example, if $H$ is positive-definite, then  $N_0=|\mathrm{Tr}H|$ is guaranteed to normalize the spectrum to a value between $0$ and $1$, and $T_0=0$ is a suitable value. Similarly, if $H$ is negative-definite, the same value of $N_0$ is valid, while $T_0=1$ will move the spectrum to the positive domain for the first iteration, making the new $\Gb$ positive-definite. We use $N_0=|\mathrm{Tr}H|$, $T_0=0$, $N=1$ and $T=0$.

        The full routine starts by computing the HF Hamiltonian $H_f$, followed by application of the Jordan-Wigner transformation $\mathrm{JW}$ to obtain the qubit Hamiltonian $H$, rescaling of $H$ to the positive-definite domain using $N_0$ and $T_0$, converting it into a \texttt{scipy} sparse matrix and splitting it into four blocks of the same size (Eq. \ref{e:hamp2}), then inserting them into Eq. \ref{broot0}, and recursively applying Eq. \ref{broot0rec} ($\mathrm{SBD}$ in Eq. \ref{eq:routine2}) for a choice of compression path $\alpha\oplus\nu\oplus\cdots$ down to a desired matrix size. An example of this routine with two compression steps and recovery of the original spectrum $\gamma$ is given in Eqs.  \ref{eq:routine1} through  \ref{eq:routinelast}. If $\mathrm{Eigensolve}$  is a VQE routine, then $\gamma$ is a single eigenvalue, the ground state. The same equations are valid for the full spectrum.
        
        \begin{align} \label{eq:routine1}
                H &= \mathrm{JW}(H_f)
            \\ H^{\prime} &= \frac{1}{N_0}H + T_0I
            \\ \label{eq:routine2}  \Gb_\alpha &= \mathrm{SBD}(H^{\prime},\, \alpha)
            \\ \Gb_\alpha^{\prime\prime} &= \frac{1}{N} \Gb_{\alpha}\Gb_{\alpha}^\dagger + TI
            \\  \Gb_{\alpha\oplus\nu} &= \mathrm{SBD}(\Gb_{\alpha}^{\prime\prime},\,\nu) \label{eq:sbdmid}
            \\ \Gb_{\alpha\oplus\nu}^{\prime\prime} &= \frac{1}{N} \Gb_{\alpha\oplus\nu}\Gb_{\alpha\oplus\nu}^\dagger + TI
         \\   \epsilon_0 &= |\mathrm{Eigensolve}(\Gb_{\alpha\oplus\nu}^{\prime\prime})|
         \\ \epsilon_1 &= \left|\sqrt{(\epsilon_0-T)N}\right|
         \\ \epsilon_2 &= \left|\sqrt{(\epsilon_1-T)N}\right|
         \\ \gamma(2) &= (\epsilon_2-T_0) N_0  \label{eq:routinelast}
        \end{align}

\subsection{Simultaneous Block-Diagonalizability Analysis}

Matrices that lie outside the constraints imposed over $H$ for the SBD algorithm will have an additional error associated with non-commutativity. As a direct consequence of the layout of the algorithm, the values of $T$ and $N$ at each step can be optimized to minimize the error for different classes of matrices.

A set of simultaneously diagonalizable blocks is guaranteed to minimize the error for the leading block \cite{higham2001solving}, since the blocks will operate eigenvector-wise. To estimate how distant a matrix is from having its blocks simultaneously diagonalizable, but without eigensolving the blocks, we exploit the fact that two diagonalizable matrices that commute are simultaneously diagonalizable \cite{gantmakher2000theory}, and compute an adapted Frobenius norm $N(H)$, the square root of the projective sum of entry-wise-squared commutators (Eq. \ref{eq:norm}).

\begin{align} \label{eq:norm}
    N(H) = \sqrt{ \sum_{q^\prime=0}^{n}\sum_{q=0}^{n}\langle q| \left( [\Ab, \Db]^{(2)} + [\Bb, \Cb]^{(2)} + [\Ab, \Cb]^{(2)} \right) | q^\prime\rangle}
\end{align}

\begin{figure}[ht]
    \centering
    \includegraphics[width=1\linewidth]{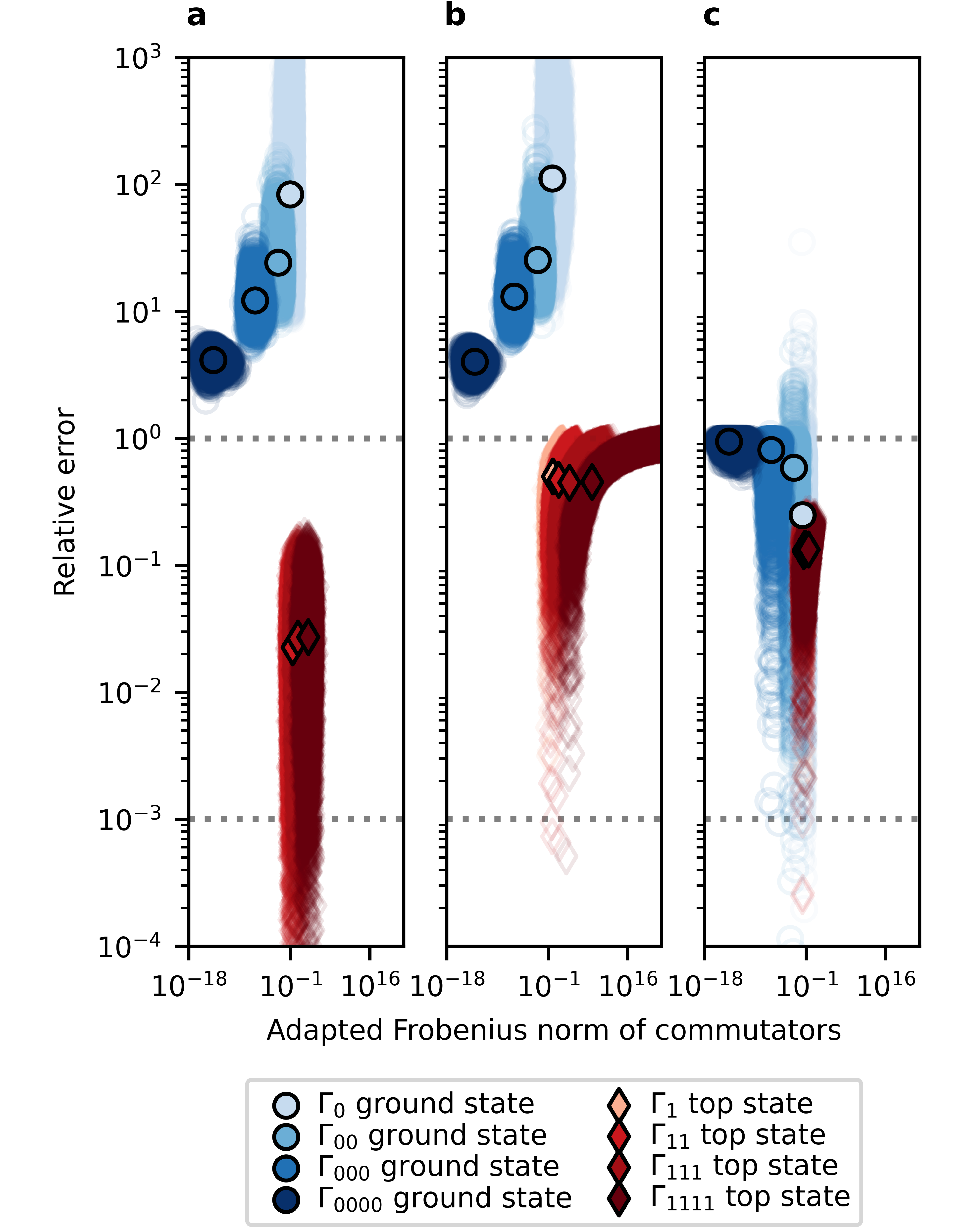}
    \caption{Relative error versus adapted Frobenius norm for ground state (circles) and top state (diamonds) estimation after $1$, $2$, $3$ and $4$ steps of SBD compression, for SD (a), PSD (b) and G (c) populations.}
    \label{fig:y}
\end{figure}

This metric allows us to analyze the relation between error, compression, and commutativity (as an indicator of simultaneous diagonalizability), so that the Hamiltonian can be categorized, and it can be decided whether it is advantageous or not to proceed with the compression.

We compute $N(H)$ and the relative error for three sets of random Hermitian matrices: (i) with simultaneously diagonalizable blocks (SD population), (ii) with simultaneously diagonalizable blocks with a small unitary perturbation (PSD population), and (iii) with random Hermitian matrices that may or may not commute (G population).

\subsection{Efficiency Analysis}

    For arbitrary matrices that do not have dimensions divisible by $2$, the matrix should be padded (extended with diagonal elements) to an even dimension. If padded to the closest dimension to $2^n$, with $n$ an integer number, the matrix will be maximally compressible. Padding is not needed for qubit Hamiltonians, as any qubitification routine will already map a Fermionic Hamiltonian or other Hermitian matrix into a matrix of size $(2^n,2^n)$.

    In summary, $k$ applications of the SBD procedure for each block-eigenvalue halve the dimension of the target $2^n$-sized matrix  $k$ times, resulting in a compression $C(k)$ that grows with a decelerated power law from $50\%$ (single compression) towards $100\%$ (full compression) in matrix size (Eq. \ref{e:compress}). 
        \begin{align} \label{e:compress}
            C(k) &= \left(1-\frac{2^{n-k}}{2^n} \right)\cdot 100\%  \\ \notag &= (1-2^{-k})\cdot 100\% 
        \end{align}

        \begin{figure}
            \centering
        \includegraphics[width=1\linewidth]{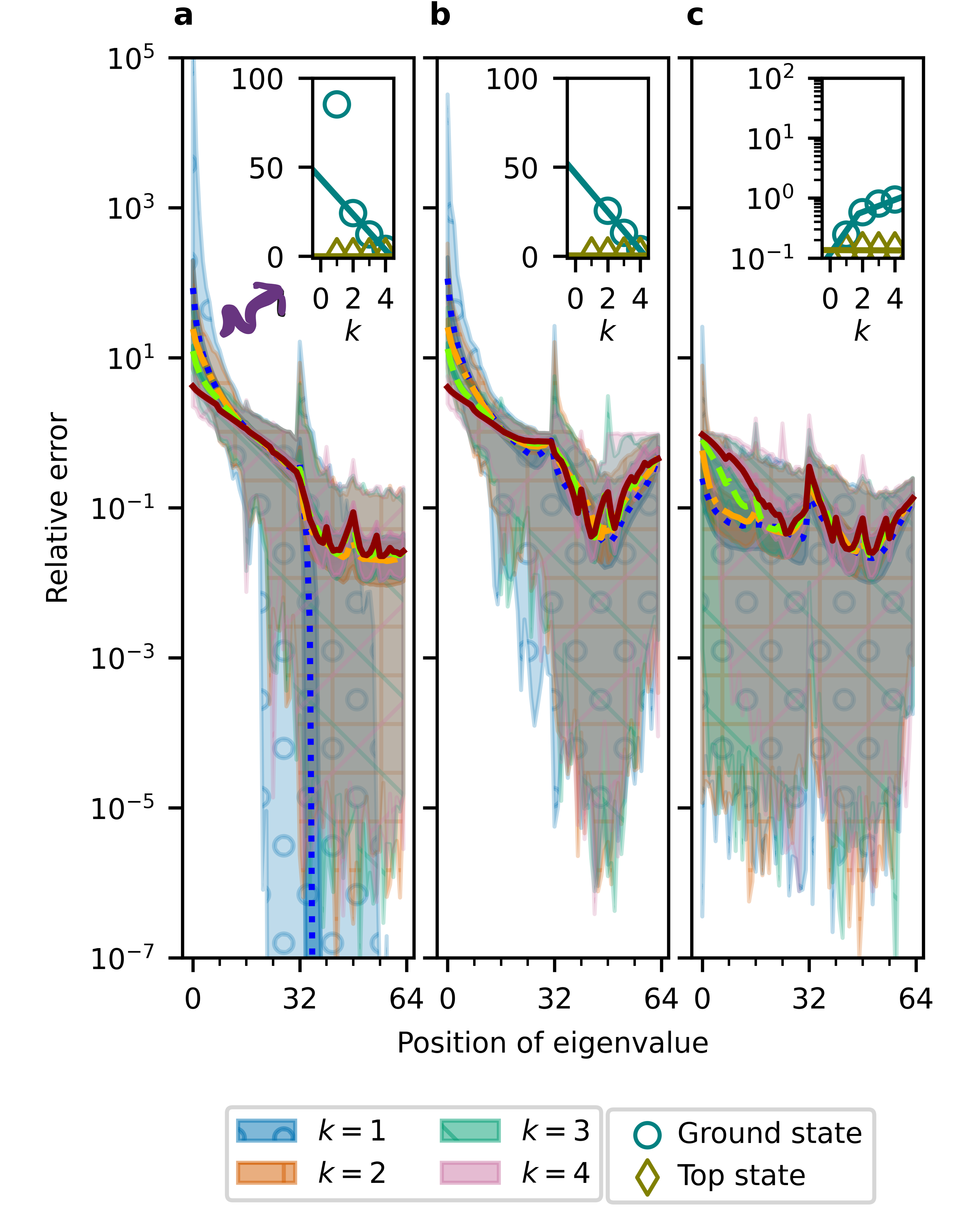}
            \caption{Relative error profiles for the full spectra after four steps of SBD compression for 10000 random Hermitian matrices of size $(2^6,2^6)$, from (a) simultaneously diagonalizable origin, (b) simultaneously diagonalizable origin with small perturbation and (c) general, non-commutative origin.}
            \label{fig:x}
        \end{figure}

    Solving Eq. \ref{e:compress} for $k(C)$, one obtains the number of SBD compressions $k$ needed to reduce the matrix size by at least a given percentage $C$, valid for any matrix size (Eq. \ref{e:thek}). Since $k$ must be a natural number, we use the ceiling notation to round the result to the closest superior integer.

        \begin{align} \label{e:thek}
            k(C) = \left\lceil -\log_2\left(1-\frac{C}{100\%} \right) \right\rceil
        \end{align}
        As an usage example of Eq. \ref{e:thek}, a compression of $90\%$ in matrix size requires just four applications of the SBD algorithm, while a $99\%$ compression requires seven applications, independently of the original matrix size.

    \begin{figure}[ht]
        \centering
        \includegraphics[width=0.9\linewidth]{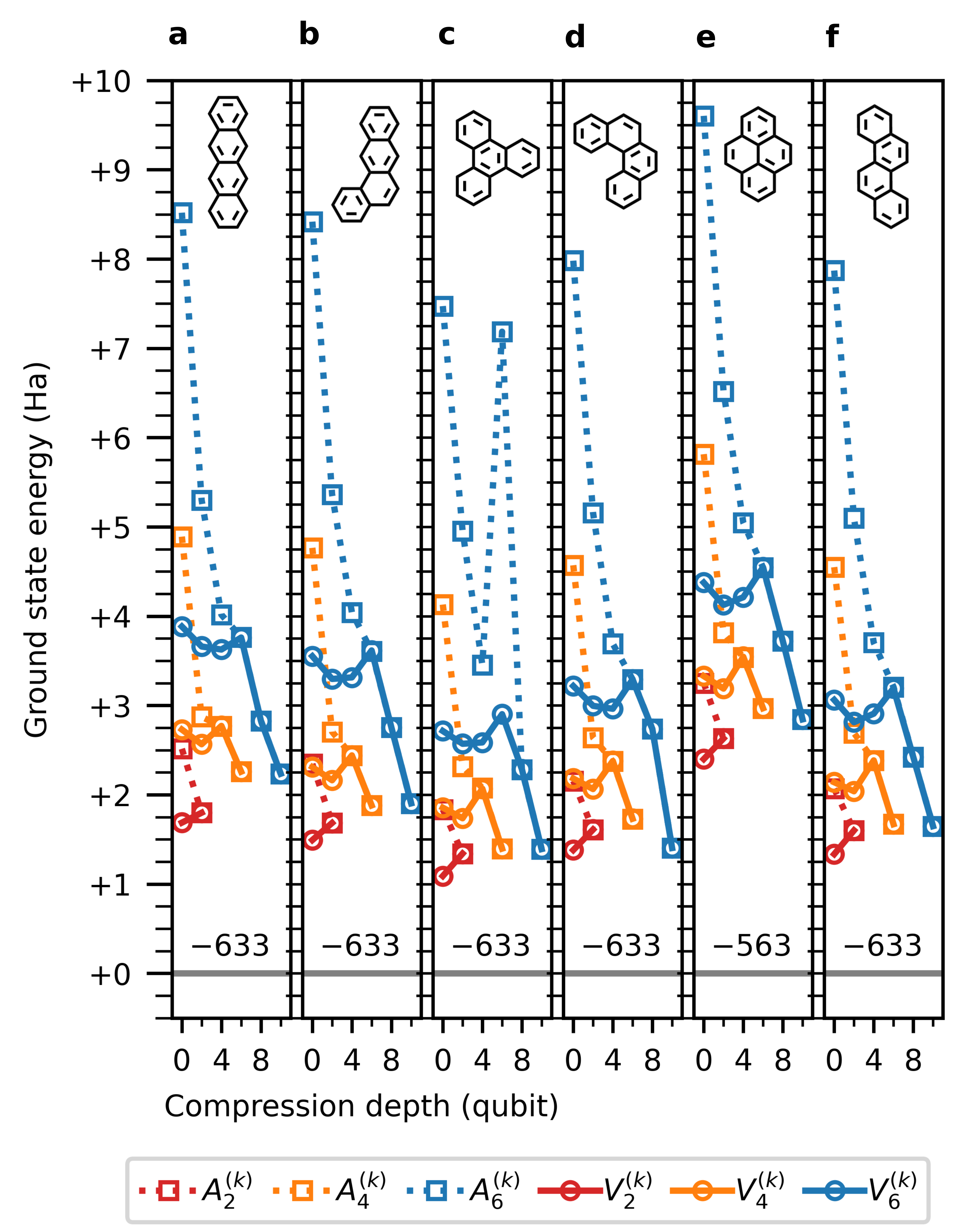}
        \caption{\textbf{Ground state energies as functions of the Hamiltonian size for three sets of electronic active space, using Arnoldi eigensolver and VQE eigensolver.} Compression of up to half the number of qubits keeps the VQE ground state slightly stable for any active space.}
        \label{f:eigenvalues}
    \end{figure}

        The original matrix size only influences the compression rate when the matrix requires padding. In the worst case, its dimension $s$ is one unit above an integer power on base $2$, i.e. $s=2^{n-1}+1$, and the matrix is padded to the next integer power $2^n=2^{\lceil\log_2s\rceil }$, before being compressed to the size $2^{n-k}$, resulting in Eq. \ref{eq:paddedcompression}. In this expression, it is visible that the matrix expands by padding when $k=0$ before starting the exponentially deep compression when $k>0$.
        \begin{align} \label{eq:paddedcompression}
            C^\prime(s,k) = \left( 1- \frac{2^{\lceil\log_2s\rceil-k}}{s} \right)\cdot 100\%
        \end{align}

        The ground states are calculated using (i) the Arnoldi method \cite{lehoucq1998arpack} provided by \texttt{scipy.sparse.linalg.eigs} and (ii) the Variational Quantum Eigensolver routine\cite{peruzzo2014variational} provided by the \texttt{pennylane} \cite{bergholm2018pennylane} python package. From them, the absolute error and speed performance are computed and plotted for analysis.

        The results are finalized with the plot of the molecular energy ranking, the problem of sorting molecules by ground state to decide which one is more stable. This plot takes the full list of twenty four models on the horizontal axis and the energy rank on the vertical axis. The models are labeled as $A_s^{(k)}$ for the Arnoldi method with active space $(s,s)$ and compression depth $k$, and as $V_s^{(k)}$ for the VQE method with respective $s$. The standard (uncompressed) model for a given active space $s$ is defined by $k=0$.
        
        Tiles of the molecular structures are placed at their respective orders of magnitude. From this graph, we compute the probability of success of using an SBD-VQE model $V_s^{(k)}$ relative to the uncompressed VQE, $V_s^{(0)}$, as the number of patterns that matches the uncompressed VQE pattern divided by the total number of SBD-VQE patterns.

    \begin{figure}[ht]
        \centering
        \includegraphics[width=0.9\linewidth]{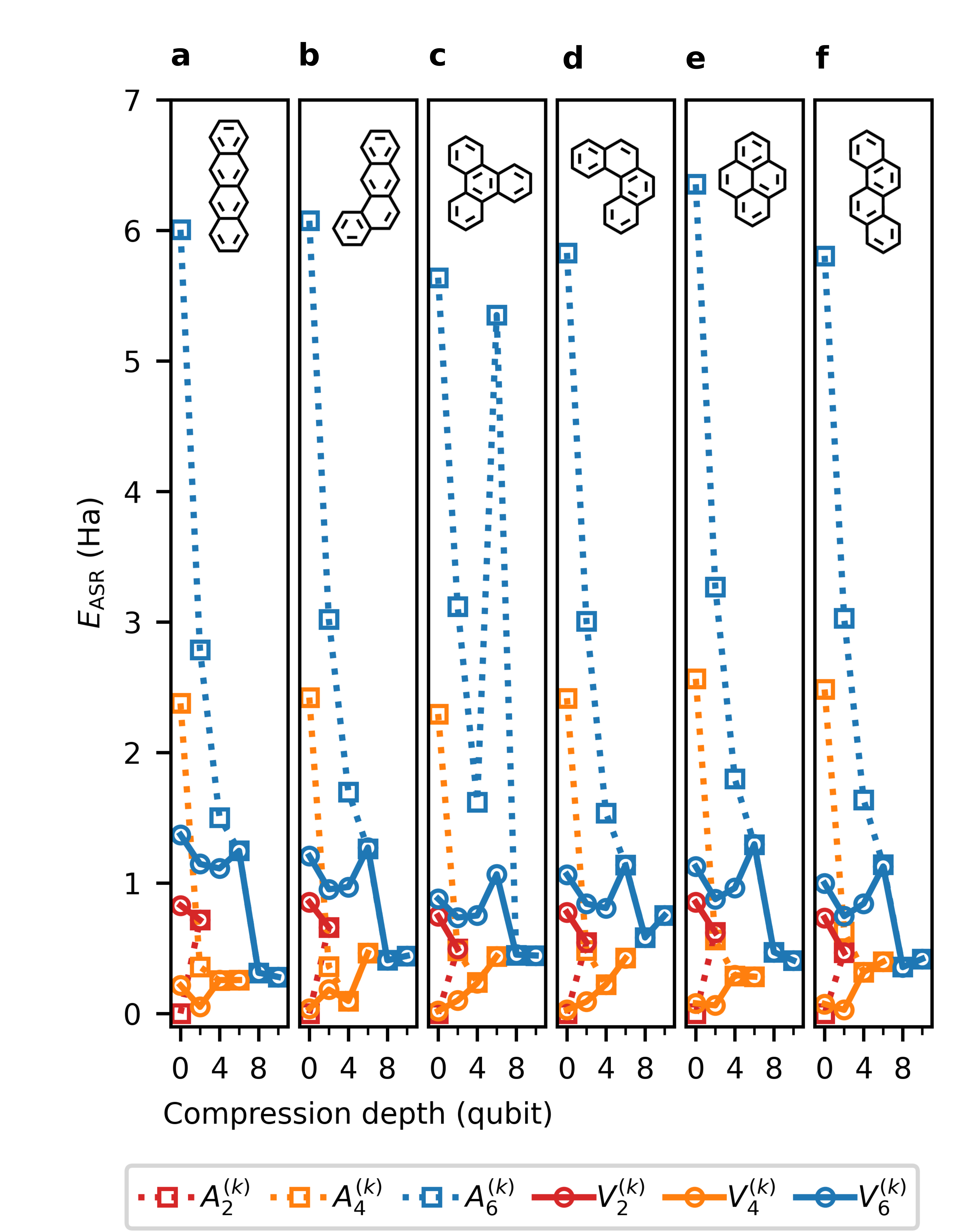}
        \caption{\textbf{Absolute error of SBD as active space reduction for the ground state energies as function of the compression depth in qubits for three sets of active spaces, using Arnoldi eigensolver and VQE eigensolver.}}
        \label{f:errors}
    \end{figure}
    
\section{Results}
    In this section, we analyze the compression rate, eigenvalues, errors, speed and molecular energy ranking of the application of the SBD compression and subsequent eigensolving of the Hartree-Fock Hamiltonians in STO-3G basis of tetracene, benz(a)anthracene, triphenylene, benzo[c]phenanthracene, pyrene and chrysene, respectively, using the Arnoldi method from \texttt{scipy} and the VQE simulation method from \texttt{pennylane}.

    The error performance of the SBD algorithm was studied for three categories of random matrices. In Figs. \ref{fig:y} and \ref{fig:x}, the first plot shows the results for compressed Hermitian matrices originated from a set of four simultaneously diagonalizable blocks ($\Ab, \Bb, \Cb$, and $\Db$); the second plot contains the results for simultaneously diagonalizable blocks with a small perturbation; the third plot contains the results for general non-commutative Hermitian matrices.

   In Fig. \ref{fig:y}, the circles (ground blocks) and the diamonds (top blocks) grow from lighter to darker along with the compression depth, with the median of the distribution marked by a black-edged circle. The monomial fit of $E_\mathrm{rel} = 40.543 N^{0.077457} + 1.1184$ for the ground blocks of Fig. \ref{fig:y}-a and of $E_\mathrm{rel} = 40.013 N^{0.062546} - 1.3025$ for those of Fig. \ref{fig:y}-b show that the blocks of $\Gb_0$ had the highest Frobenius norm of commutators $N$ and also the highest errors $E_\mathrm{rel}$, thus, following the expected proportionality between the norm of commutators and the relative error. In contrast, $\Gb_0$ in Fig \ref{fig:y}-c shows an inverse relation, with the relative error increasing as the Frobenius norm of commutators decreases.

    The relative error for the spectra of $6$-qubit-large random Hermitian matrices is shown in Fig. \ref{fig:x}, where the ground states show the largest median errors (central lines), with a descending, fluctuating slope towards the top state. For the set of simultaneously diagonalizable parents (Fig. \ref{fig:x}-a), the first compression leads to the top block having nearly zero error, while error is introduced in the subsequent compression steps to a range of medians between $1\%$ and $10\%$.

    The descendants of perturbed simultaneously diagonalizable matrices (center) show a very similar behavior, except that the error for the top block after the first compression is no longer null, with medians located between  $10\%$ and $100\%$.

    The set of non-simultaneously-diagonalizable parents (Fig \ref{fig:x}-c) shows smaller error for both the ground and the top states, in comparison with the perturbed set. However, with median still ranging from $10\%$ to $100\%$ for both states.

    In all three plots, the relative error distributions are skewed towards the smaller magnitude sector, covering the range from $0.001\%$ to $100\%$ for the top state. Most importantly, the ground state error reduces with increasing compression depth in the first and second cases.
   
    The insets of Fig. \ref{fig:x} show a linear tendency for the median relative error as a function of the compression depth for the ground states (circles) and the top states (diamonds). The steep curves for the ground states in the SD population, the PSD population and the G population are described by the least-squares linear fit in Eqs. \ref{eq:errorvsk0}, \ref{eq:errorvsk1} and \ref{eq:errorvsk2}, respectively. Their linear tendencies are quantified by the Pearson linear correlation coefficients of $r_\mathrm{SD}=-0.99381$, $r_\mathrm{PSD}=-0.99700$,  $r_\mathrm{G}=0.98724$, with five significant digits.

    \begin{align} 
        \label{eq:errorvsk0} E_\mathrm{rel,SD}(k)  &= -10.059 k + 43.700,
        \\ \label{eq:errorvsk1} E_\mathrm{rel,PSD}(k) &= -10.582 k + 45.885,
        \\ \label{eq:errorvsk2} E_\mathrm{rel,G}(k) &=  0.17650 k + 0.13649.
    \end{align}
    
    The ground blocks and the top blocks behave very differently, as the median relative errors for the top blocks show nearly absence of linear correlation with the compression depth.

    The SBD routine reduces the magnitude of ground states of the Arnoldi method almost linearly, as seen in Fig. \ref{f:eigenvalues}. In contrast, the SBD-VQE method shows a convex curve between the original Hamiltonian and its compressed version to half the number of qubits for all active spaces, except for the $(2,2)$ one. At half the number of qubits, the linear tendency towards the Arnoldi method becomes prevalent. An outlier is seen in Fig. \ref{f:eigenvalues}-c, likely attributed to the intrinsinc approximation errors of the compression. Fig. \ref{f:eigenvalues} shows that the SBD method has an effect similar to active space reduction, with the most compressed blocks approaching the results of the $A_2^{(0)}$ and $V_2^{(0)}$ models.

    \begin{figure}[ht]
        \centering
        \includegraphics[width=1\linewidth]{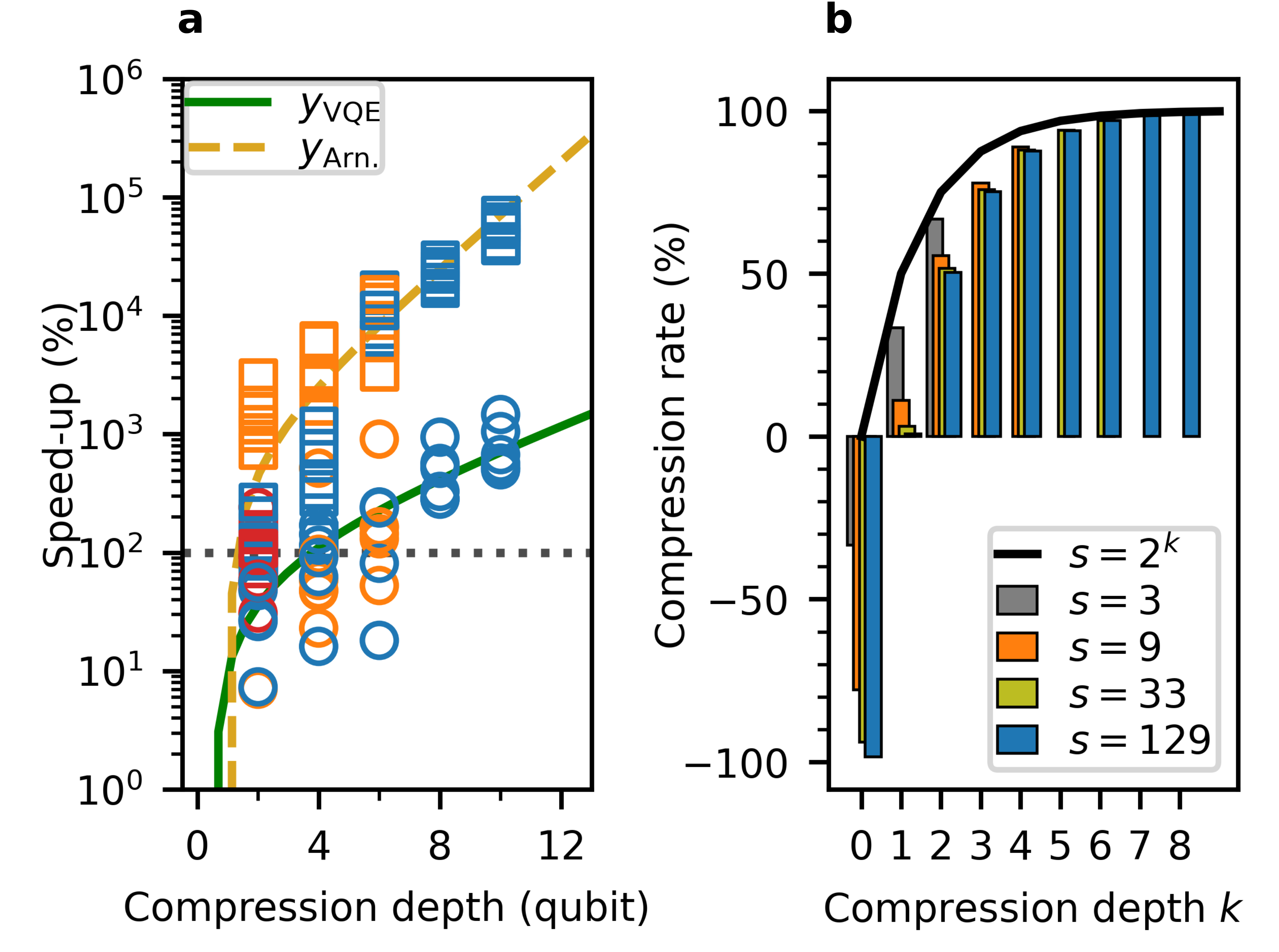}
        \caption{\textbf{Relative speed and theoretical compression rate. (a) Speed ($y$) of the SBD-VQE and SBD-Arnoldi eigensolvers relative to the speed of the uncompressed methods for each active space ($V_{6}^{(0)}$) for each molecule, expressed as function of the compression depth $k$. Markers above the dashed line indicate advantage. Here, compression depth grows from right to left. Marker colors are the same as in Fig. \ref{f:eigenvalues}. (b) Compression rate as a function of the number of compressed qubits $k$ for different original matrix sizes $s$. The negative compression rates at $k=0$ are due to padding prior to application of SBD.
        }}
        \label{fig:rate} 
    \end{figure}
    The absolute errors of SBD as active space reduction were computed with respect to the ground states of the $A_2^{(0)}$ model (Fig. \ref{f:errors}), as they are closest to the reference values of the two molecules present at CCCBDB \cite{NIST_CCCBDB} for more sophisticated basis sets, tetracene (Fig. \ref{f:errors}-a) and pyrene (Fig. \ref{f:errors}-e). As the aim of this study is to compare eigensolvers with and without compression, we do not use the exact ground state values of these molecules as merit figures.

    The graphs in Fig. \ref{f:errors} show that the error against $A_2^{(0)}$ decreases with the number of compressions for the largest active space. The Arnoldi method shows a larger descending slope up to half the number of qubits, where both VQE and the Arnoldi methods start to converge to a common line. Furthermore, the convex curve of the VQE at half the number of qubits is an indication that the SBD compression shows a stable result for the first $n/2$ compressions, where $n$ is the Hamiltonian size in number of qubits.

    The advantage of using such a powerful compression algorithm is shown in Fig. \ref{fig:rate}-a, where square markers represent the Arnoldi method and circles represent the VQE method. After calculation of an exponential fit, it is evident that the speed-up grows exponentially as the compression depth increases. The fitting curves in Eq.\ref{eq:arnspeed} and Eq. \ref{eq:vqespeed} show that the SBD routine speeds up the VQE simulation exponentially.

    \begin{figure*}[ht]
        \centering
        \includegraphics[width=1\linewidth]{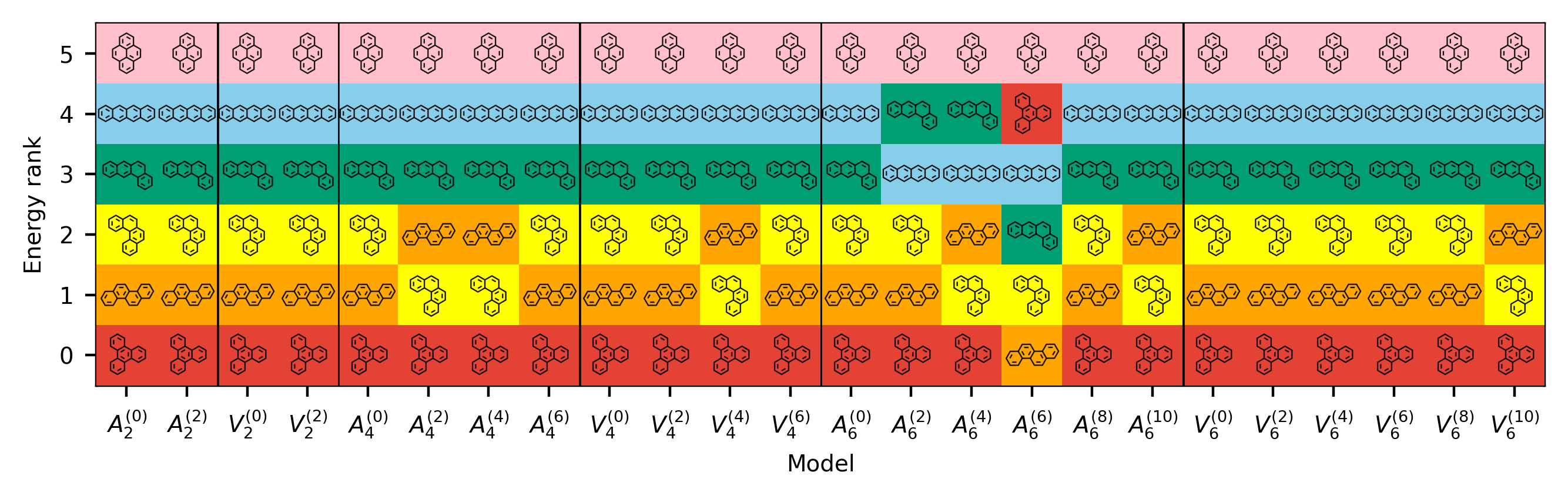}
        \caption{\textbf{Molecular energy ranking for the six studied PAHs for each model for different compression depths.} The horizontal axis shows all the twenty four models used in this study, indicating the eigensolver, active space (number of active electrons, number of active orbitals) and compression depth in qubits, respectively. The colors are merely to enhance pattern visualization.}
        \label{f:sorted}
    \end{figure*}

\begin{align} \label{eq:vqespeed}
    y_\mathrm{VQE}(k) &=  80.33\exp(0.23k) -91.10,
\\ \label{eq:arnspeed} 
    y_\mathrm{Arn.}(k) &= 395.26\exp(0.52k) -670.87.
\end{align}

    In case the matrix size is an odd number, Fig. \ref{fig:rate}-b shows that padded matrices reach compression rates similar to qubitified matrices after a few compression steps. The black curve is a plot of Eq. \ref{e:compress}, while the bars are Eq. \ref{eq:paddedcompression} for different original matrix sizes $s$, starting at the padded size (negative compression). If the matrix size for the compression rate is expressed in number of qubits, then the compression rate grows linearly with compression depth, with each step reducing one qubit.

    As a direct application of the SBD algorithm, Fig. \ref{f:sorted} shows the ground states of the six TAHs ranked by ascending magnitude from bottom to top, for each combination of active space, eigensolver and compression depth. The figure shows that 77.8\% of the SBD-assisted VQE will fully match the ground state ranking of the uncompressed VQE for an arbitrary choice of active spaces from among $(2,2), (4,4)$, and $(6,6)$. This result is more than double the value for the SBD-Arnoldi method ($33.3\%$), indicating that the compression method is better suited for VQE than for the Arnoldi method. Increasing the size of the active space worsened the success rate of the Arnoldi eigensolver, since $P_\mathrm{U}$ of VQE is higher than that of Arnoldi and $P_\mathrm{R2}$ is $83.3\%$, higher than the $50\%$ of $P_\mathrm{PR1}$, as summarized in Tab. \ref{t:probs}. 

    As consequence, the molecular energy ranking shows strong agreement between SBD-VQE for several compression depths and VQE alone, in addition to the speed-up it provides.

    \begin{table}[ht]    \small
    \caption{Probabilities that an arbitrary model in the SBD-assisted molecular energy ranking represents (i) an uncompressed model ($P_\mathrm{U}$), (ii) SBD as an active space reduction ($P_\mathrm{ASR}$), (iii) VQE as Arnoldi eigensolver for first half of compressions ($P_\mathrm{R1}$), (iv) VQE as Arnoldi eigensolver for second half of compressions ($P_\mathrm{R2}$), based on Fig. \ref{f:sorted}, followed by the average relative error of SBD as active space reduction $\langle E_\mathrm{ASR} \rangle$ for all samples, and the median speed-up $\tilde{y}$ for the largest active space.\label{t:probs}}
    \begin{tabular}{lllllll}
    \hline
    Eigensolver & $P_\mathrm{U}$ & $P_\mathrm{ASR}$ & $P_\mathrm{R1}$ & $P_\mathrm{R2}$  & $\langle E_\mathrm{ASR} \rangle$&$\tilde{y}$\\ \hline
    SBD-Arnoldi     & $33.3\%$& $40\%$& -               & -                & $0.23\%$& $5945.8\%$\\
    SBD-VQE         & $77.8\%$& $80\%$& $50\%$& $83.3\%$& $0.09\%$& $164.16\%$\\ \hline
    \end{tabular}
    \end{table}
    
    \section{Discussion} \label{s:appl}
        In this section, we discuss the implications and limitations of the use of SBD for energy ranking, in addition to motivating applications to other physical problems.

        In lambda matrix polynomials of dynamical systems, the coefficients $A_0$, $A_1$ and $A_2$ represent the stiffness, the damping coefficient and the mass of the system, respectively \cite{tisseur2001quadratic}. By analogy, the conditions for a low-error SBD discussed in section \ref{sec:intro} are common physical approximations of the mass as a unit and isotropic, and of $\mathrm{Tr_{block}}H$ as a small generalized damping.
        
        The geometry optimization problem is a special case of molecular energy ranking where an optimization loop minimizes the energy of candidate geometries to find the most stable geometry \cite{delgado2021variational}. The probability of $77.8\%$ of reproducing the uncompressed VQE in the energy ranking of the six PAHs indicates that it is a strong candidate to tackle geometry optimization and other energy optimization and classification problems using VQE. 

        In the set of ranked TAHs, triphenylene has the smallest rank, followed by chrysene, benzo[c]phenanthracene, benz(a)anthracene, tetracene and pyrene, indicating that pyrene is the least likely to occur as a subproduct of a reaction network, where it competes with the other five TAHs. 

        The error, speed and active space reduction analogy qualify the SBD algorithm for four immediate applications: symbolic expansion of the diagonalization routine, vector compression, principal component estimation in PCA, and Ansatz optimization for VQE.
        
        As the first application, SBD allows for a symbolic expression of the eigenvalues since it relies on matrix addition, multiplication and the Newton-Schulz expansion for matrix square roots. This expansion with the use of partial inversion\cite{lima2024unitarization} to compute the inverse matrix is the main factor that makes the SBD compression symbolically treatable in a semi-analytical way.

    As the second application, the eigenvalue decomposition theorem allows an extension of the SBD compression to include vector compression, thus making the compression of Ansatze (the guess initial states) in VQE possible. Rewriting an arbitrary vector $|\psi\rangle\in \mathbb{C}$ as a matrix $\rho = |\psi\rangle\langle\psi|$, the Spectral Theory \cite{hassani2013mathematical} states that $\rho$ has only one non-null eigenvalue. If $|\psi\rangle$ is normalized, then this eigenvalue equals $1$. As consequence, after SBD compression into $\rho_{[1]}$, the eigenvector that corresponds to this eigenvalue will always be located in the upper-magnitude block-eigenvalue, and will be far enough from the rest of the spectrum to be distinguishable even when accounting for decimal noise.

    Therefore, after SBD compression (Eqs. \ref{e:density, e:density2} ), an eigenvector search in the upper-magnitude block-eigenvalue around the eigenvalue $1$ will find the estimated compressed vector. Subsequently, the vector may be normalized on its norm, then multiplied by the norm of the uncompressed vector to recover the desired vector.
        
        \begin{align} \label{e:density}
                \rho &= |\psi\rangle \langle \psi|
                \\ \label{e:density2} \rho |\psi\rangle &= |\psi\rangle \langle \psi  |\psi\rangle = 1 |\psi\rangle
        \end{align}
        
    As shown in Eq. \ref{e:density3}, the compressed matrix loses unitarity and has $N-1$ eigenvalues $a_i\approx 0$, with respective eigenvectors $|\phi_i\rangle$ due to approximation errors. The subsequent step is the application of a sparse eigensolver to approximate the eigenvector $|\psi_{[1]}\rangle$ of $\rho_{[1]}$ around the eigenvalue $1$, obtaining the eigenvalue $b\approx 1$.

        \begin{align} \label{e:density3}
                \rho_{[1]} &=b|\psi_{[1]}\rangle \langle \psi_{[1]}| + \sum_{i=1}^{N-1} a_i|\phi_i\rangle\langle\phi_i|
                \\ \rho_{[1]} |\psi_{[1]}\rangle &=b |\psi_{[1]}\rangle \langle \psi_{[1]}  |\psi_{[1]}\rangle =  b|\psi_{[1]}\rangle
        \end{align}

        As the third application, the ground-state problem can be transformed into the leading eigenvalue problem by simple multiplication of the original matrix by $-1$, followed by adding an identity matrix with a constant coefficient. These transformations are already used in the SBD algorithm to move the spectrum into a stable domain prior to eigensolving.
        
        This transformation is of special importance in PCA for the computation of the eigenvector of the largest eigenvalue of the covariance matrix of a dataset in order to project the dataset into a relevant subspace \cite{greenacre2022principal}. Therefore, both the SBD algorithm and its defended application of eigenvector compression are expected to extend the use of SBD-VQE to perform a PCA or, at least, approximate the largest eigenvalue and eigenvector to reduce their search space in other algorithms.

        The fourth application is Ansatz optimization, due to the high speed of the SBD routine. The fidelity and speed of VQE methods are highly dependent on the quality of the initial guess state (Ansatz) \cite{zhang2022variational}, making the error-speed exchange profile viable to find an eigenvector closer to the actual solution prior to the final application of a VQE to estimate a more accurate solution.

        These applications are objects of future research with vast impacts in electronic structure simulation, geometry optimization, and machine learning. 
        
    \section{Conclusion}
        In this paper, we generalized the Sridhara method to matrix secular equations and studied its error profile against compression depth and block commutativity, in addition to testing its efficiency to rank six TAHs of relevance in the Dry Reforming of Methane and graphene synthesis.
        
        The 77.8\% probability of reproducing the energy ranking of uncompressed Hamiltonians, in addition to the elevated median speed-up of $164.16\%$, and small average error of $0.09\%$ make SBD-VQE a better combination than SBD-Arnoldi for ground state estimation, even though the Arnoldi method showed higher speed-up after compression. 
        
        We conclude that the SBD algorithm is a fast and simple block eigensolver with a small relative error when its constraints are flexibilized.  Future steps should further investigate both a symbolic procedure based on the present results and the potential applications of the SBD algorithm for the study of more complex molecules.

\section*{Author contributions}
Dennis Lima: Conceptualization, Data curation, Investigation, Visualization, Writing – original draft, Writing – review and editing. Saif Al-Kuwari: Funding acquisition, Supervision, Writing – review and editing.

\section*{Conflicts of interest}
There are no conflicts to declare.

\section*{Data availability}
The code for the Sridhara Block Diagonalization can be found at \href{https://www.github.com/partialg}{https://www.github.com/partialg} with an interactive notebook.

\section*{Acknowledgements}
This project was funded by Qatar Center for Quantum Computing and Hamad Bin Khalifa University. The funder played no role in study design, data collection, analysis and interpretation of data, or the writing of this manuscript.

\bibliography{main}

@article{gemeinhardt2023quantum,
author = {Gemeinhardt, Felix and Garmendia, Antonio and Wimmer, Manuel and Weder, Benjamin and Leymann, Frank},
title = {Quantum Combinatorial Optimization in the NISQ Era: A Systematic Mapping Study},
year = {2023},
issue_date = {March 2024},
publisher = {Association for Computing Machinery},
address = {New York, NY, USA},
volume = {56},
number = {3},
issn = {0360-0300},
url = {https://doi.org/10.1145/3620668},
doi = {10.1145/3620668},
journal = {ACM Comput. Surv.},
articleno = {70},
pages    = {0},
numpages = {36}
}

@techreport{dennis1971matrix,
  title     = {On the Matrix Polynomial, Lambda-matrix and Block Eigenvalue Problems},
  author    = {Dennis Jr., John E. and Traub, J.F. and Weber, R.P.},
  number    = {71},
  year      = {1971},
  institution = {Cornell University},
  URL       = {https://ecommons.cornell.edu/items/0a874c92-6f02-4a20-8dc0-81530f914cb5/full}
}

@article{zhang2001lanczos,
  title={Lanczos subspace filter diagonalization: Homogeneous recursive filtering and a low-storage method for the calculation of matrix elements},
  author={Zhang, Hong and Smith, Sean C},
  journal={Physical Chemistry Chemical Physics},
  volume={3},
  number={12},
  pages={2282--2288},
  year={2001},
  publisher={Royal Society of Chemistry}
}

@article{meyer2025solving,
  title={Solving Linear DSGE Models with Bernoulli Iterations: A. Meyer-Gohde},
  author={Meyer-Gohde, Alexander},
  journal={Computational economics},
  volume={66},
  number={1},
  pages={593--643},
  year={2025},
  publisher={Springer},
  doi={https://doi.org/10.1016/j.econmod.2024.106670}
}

@book{gohberg1982matrix,
author = {Gohberg, I. and Lancaster, P. and Rodman, L.},
title = {Matrix Polynomials},
publisher = {Academic Press},
year = {1982},
doi = {10.1137/1.9780898719024},

}

@inbook{Guicciardini_2016, 
    place    ={Cambridge}, 
    series   ={Cambridge Companions to Philosophy}, 
    title    ={A brief introduction to the mathematical work of Isaac Newton}, 
    booktitle={The Cambridge Companion to Newton}, 
    publisher={Cambridge University Press}, 
    author   ={Guicciardini, Niccolò}, 
    editor   ={Iliffe, Rob and Smith, George E.Editors}, 
    year     ={2016}, 
    pages    ={382–420}, 
    collection={Cambridge Companions to Philosophy}
}

@article{tjalling1995,
 ISSN = {00361445, 10957200},
 URL = {http://www.jstor.org/stable/2132904},
 author = {Tjalling J. Ypma},
 journal = {SIAM Review},
 number = {4},
 pages = {531--551},
 publisher = {Society for Industrial and Applied Mathematics},
 title = {Historical Development of the Newton-Raphson Method},
 urldate = {2025-11-27},
 volume = {37},
 year = {1995}
}

@article{Bernoulli1728,
  author    = {Daniel Bernoulli},
  title     = {Observationes de seriebus recurrentibus},
  journal   = {Commentarii Academiae Scientiarum Imperialis Petropolitanae},
  volume    = {3},
  year      = {1728},
  note      = {Published 1732},
  pages     = {85--100}
}

@article{dutka1995early,
  title={On the early history of Bessel functions},
  author={Dutka, Jacques},
  journal={Archive for history of exact sciences},
  pages={105--134},
  year={1995},
  publisher={JSTOR}
}

@article{traub1966class,
  title={A class of globally convergent iteration functions for the solution of polynomial equations},
  author={Traub, Joseph F},
  journal={Mathematics of Computation},
  volume={20},
  number={93},
  pages={113--138},
  year={1966},
  publisher={JSTOR},
doi={https://doi.org/10.2307/2004275}
}

@article{davis1981numerical,
author = {Davis, George J.},
title = {Numerical Solution of a Quadratic Matrix Equation},
journal = {SIAM Journal on Scientific and Statistical Computing},
volume = {2},
number = {2},
pages = {164-175},
year = {1981},
doi = {10.1137/0902014},
URL = { https://doi.org/10.1137/0902014 },
eprint = { https://doi.org/10.1137/0902014}
}

@article{davis1983algorithm,
author = {Davis, George J.},
title = {Algorithm 598: an algorithm to compute solvent of the matrix equation {$AX^2 + BX + C = 0$} },
year = {1983},
issue_date = {June 1983},
publisher = {Association for Computing Machinery},
address = {New York, NY, USA},
volume = {9},
number = {2},
issn = {0098-3500},
url = {https://doi.org/10.1145/357456.357463},
doi = {10.1145/357456.357463},
journal = {ACM Trans. Math. Softw.},
pages = {246–254},
numpages = {9}
}

@book{gohberg2009matrix,
author = {Gohberg, I. and Lancaster, P. and Rodman, L.},
title = {Matrix Polynomials},
publisher = {Society for Industrial and Applied Mathematics},
year = {2009},
doi = {10.1137/1.9780898719024},
address = {},
edition   = {},
URL = {https://epubs.siam.org/doi/abs/10.1137/1.9780898719024},
eprint = {https://epubs.siam.org/doi/pdf/10.1137/1.9780898719024},
note = {This SIAM edition is an unabridged republication of the work first published by Academic Press, Inc., 1982.}
}

@article{dennis1978algorithms,
  title={Algorithms for solvents of matrix polynomials},
  author={Dennis, Jr, JE and Traub, Joseph Frederick and Weber, Robert P},
  journal={SIAM Journal on Numerical Analysis},
  volume={15},
  number={3},
  pages={523--533},
  year={1978},
  publisher={SIAM}
}

@article{dennis1976algebraic,
  title={The algebraic theory of matrix polynomials},
  author={Dennis, Jr, John E and Traub, Joseph F and Weber, Robert P},
  journal={SIAM Journal on Numerical Analysis},
  volume={13},
  number={6},
  pages={831--845},
  year={1976},
  publisher={SIAM}
}

@article{lanczos1950iteration,
  title={An iteration method for the solution of the eigenvalue problem of linear differential and integral operators},
  author={Lanczos, Cornelius},
  journal={Journal of research of the National Bureau of Standards},
  volume={45},
  number={4},
  pages={255--282},
  year={1950}
}

@Article{delgado2021variational,
  author    = {Delgado, Alain and Arrazola, Juan Miguel and Jahangiri, Soran and Niu, Zeyue and Izaac, Josh and Roberts, Chase and Killoran, Nathan},
  journal   = {Physical Review A},
  title     = {Variational quantum algorithm for molecular geometry optimization},
  year      = {2021},
  number    = {5},
  pages     = {052402},
  volume    = {104},
  doi       = {10.1103/PhysRevA.104.052402},
  publisher = {APS},
  ranking   = {rank5},
}

@Book{hassani2013mathematical,
  author    = {Hassani, Sadri},
  publisher = {Springer Science \& Business Media},
  title     = {Mathematical physics: a modern introduction to its foundations},
  year      = {2013},
  doi       = {10.1007/978-3-319-01195-0},
  ranking   = {rank5},
}

@Article{naghdi2018catalytic,
  author    = {Naghdi, Samira and Rhee, Kyong Yop and Park, Soo Jin},
  journal   = {Carbon},
  title     = {A catalytic, catalyst-free, and roll-to-roll production of graphene via chemical vapor deposition: Low temperature growth},
  year      = {2018},
  pages     = {1--12},
  volume    = {127},
  doi       = {10.1016/j.carbon.2017.10.065},
  publisher = {Elsevier},
  ranking   = {rank5},
}

@Article{lu2015molecular,
  author    = {Lu, Yujie and Yang, Xiaoning},
  journal   = {Carbon},
  title     = {Molecular simulation of graphene growth by chemical deposition on nickel using polycyclic aromatic hydrocarbons},
  year      = {2015},
  pages     = {564--573},
  volume    = {81},
  doi       = {10.1016/j.carbon.2014.09.091},
  publisher = {Elsevier},
  ranking   = {rank5},
}

@Article{norinaga2007detailed,
  author    = {Norinaga, Koyo and Deutschmann, Olaf},
  journal   = {Industrial \& engineering chemistry research},
  title     = {Detailed kinetic modeling of gas-phase reactions in the chemical vapor deposition of carbon from light hydrocarbons},
  year      = {2007},
  number    = {11},
  pages     = {3547--3557},
  volume    = {46},
  doi       = {htpps://doi.org/10.1021/ie061207p},
  publisher = {ACS Publications},
  ranking   = {rank5},
}

@Article{saeed2020chemical,
  author    = {Saeed, Maryam and Alshammari, Yousef and Majeed, Shereen A and Al-Nasrallah, Eissa},
  journal   = {Molecules},
  title     = {Chemical vapour deposition of graphene—Synthesis, characterisation, and applications: A review},
  year      = {2020},
  number    = {17},
  pages     = {3856},
  volume    = {25},
  doi       = {10.3390/molecules25173856},
  publisher = {MDPI},
  ranking   = {rank5},
}

@Article{bergholm2018pennylane,
  author  = {Ville Bergholm and Josh Izaac and Maria Schuld and Christian Gogolin and Carsten Blank and Keri McKiernan and Nathan Killoran},
  journal = {arXiv},
  title   = {PennyLane: Automatic Differentiation of Hybrid Quantum-Classical Computations},
  year    = {2018},
  eprint  = {1811.04968},
  ranking = {rank5},
  url     = {https://arxiv.org/abs/1811.04968},
}

@Book{lehoucq1998arpack,
  author    = {Lehoucq, Richard B and Sorensen, Danny C and Yang, Chao},
  publisher = {SIAM},
  title     = {ARPACK users' guide: solution of large-scale eigenvalue problems with implicitly restarted Arnoldi methods},
  year      = {1998},
  doi       = {10.1137/1.9780898719628.fm},
  ranking   = {rank5},
}

@Article{peruzzo2014variational,
  author    = {Peruzzo, Alberto and McClean, Jarrod and Shadbolt, Peter and Yung, Man-Hong and Zhou, Xiao-Qi and Love, Peter J and Aspuru-Guzik, Al{\'a}n and O’brien, Jeremy L},
  journal   = {Nature communications},
  title     = {A variational eigenvalue solver on a photonic quantum processor},
  year      = {2014},
  number    = {1},
  pages     = {4213},
  volume    = {5},
  doi       = {10.1038/ncomms5213},
  publisher = {Nature Publishing Group UK London},
  ranking   = {rank5},
}

@Misc{NIST_CCCBDB,
  author       = {Russell D. Johnson III},
  howpublished = {NIST Standard Reference Database Number 101, Release 22, May 2022},
  note         = {Available at \url{http://cccbdb.nist.gov/}},
  title        = {NIST Computational Chemistry Comparison and Benchmark Database},
  year         = {2022},
  ranking      = {rank5},
}

@Article{Silvester_2000,
  author  = {Silvester, John R.},
  journal = {The Mathematical Gazette},
  title   = {Determinants of block matrices},
  year    = {2000},
  number  = {501},
  pages   = {460–467},
  volume  = {84},
  doi     = {10.2307/3620776},
  ranking = {rank5},
}

@Article{tilly2022variational,
  author    = {Tilly, Jules and Chen, Hongxiang and Cao, Shuxiang and Picozzi, Dario and Setia, Kanav and Li, Ying and Grant, Edward and Wossnig, Leonard and Rungger, Ivan and Booth, George H and others},
  journal   = {Physics Reports},
  title     = {The variational quantum eigensolver: a review of methods and best practices},
  year      = {2022},
  pages     = {1--128},
  volume    = {986},
  doi       = {10.1016/j.physrep.2022.08.003},
  publisher = {Elsevier},
  ranking   = {rank5},
}

@Article{yoon2022lossy,
  author    = {Yoon, Boram and Nguyen, Nga TT and Chang, Chia Cheng and Rrapaj, Ermal},
  journal   = {Scientific Reports},
  title     = {Lossy compression of statistical data using quantum annealer},
  year      = {2022},
  number    = {1},
  pages     = {3814},
  volume    = {12},
  doi       = {10.1038/s41598-022-07539-z},
  publisher = {Nature Publishing Group UK London},
  ranking   = {rank5},
}

@article{higham2001solving,
author = {Higham, Nicholas J. and Kim, Hyun-Min},
title = {Solving a Quadratic Matrix Equation by Newton's Method with Exact Line Searches},
journal = {SIAM Journal on Matrix Analysis and Applications},
volume = {23},
number = {2},
pages = {303-316},
year = {2001},
doi = {10.1137/S0895479899350976},
URL = {https://doi.org/10.1137/S0895479899350976
}
}

@Article{PhysRevA.106.042401,
  author    = {Yang, Xiaodong and Nie, Xinfang and Ji, Yunlan and Xin, Tao and Lu, Dawei and Li, Jun},
  journal   = {Phys. Rev. A},
  title     = {Improved quantum computing with higher-order Trotter decomposition},
  year      = {2022},
  month     = {Oct},
  pages     = {042401},
  volume    = {106},
  doi       = {10.1103/PhysRevA.106.042401},
  issue     = {4},
  numpages  = {9},
  publisher = {American Physical Society},
  ranking   = {rank5},
  url       = {https://link.aps.org/doi/10.1103/PhysRevA.106.042401},
}

@Article{greenacre2022principal,
  author    = {Greenacre, Michael and Groenen, Patrick JF and Hastie, Trevor and d’Enza, Alfonso Iodice and Markos, Angelos and Tuzhilina, Elena},
  journal   = {Nature Reviews Methods Primers},
  title     = {Principal component analysis},
  year      = {2022},
  number    = {1},
  pages     = {100},
  volume    = {2},
  doi       = {10.1038/s43586-022-00184-w},
  publisher = {Nature Publishing Group UK London},
  ranking   = {rank5},
}

@article{DAVIDSON197587,
title = {The iterative calculation of a few of the lowest eigenvalues and corresponding eigenvectors of large real-symmetric matrices},
journal = {Journal of Computational Physics},
volume = {17},
number = {1},
pages = {87-94},
year = {1975},
issn = {0021-9991},
doi = {https://doi.org/10.1016/0021-9991(75)90065-0},
url = {https://www.sciencedirect.com/science/article/pii/0021999175900650},
author = {Ernest R. Davidson}
}

@Book{sridhara1959,
  author      = {Acarya, Sridhara and Shukla, Kripa Shankar},
  editor      = {Shukla, Kripa Shankar and Ballabh, Ram},
  publisher   = {Lucknow University Dept. of Mathematics and Astronomy},
  title       = {The Patiganita of Sridharacarya, with an ancient Sanskrit commentary},
  year        = {1959},
  address     = {Lucknow, India},
  edition     = {1st},
  series      = {Hindu astronomical and mathematical texts series; no. 2},
  volume      = {1},
  chapter     = {5},
  contributor = {{S\~{S}}r\={i} Raghun\={a}tha Mandira Library (Jammu, India); Lucknow University Department of Mathematics \& Astronomy},
  language    = {Sanskrit; English},
  pages       = {68--71},
  ranking     = {rank5},
  url         = {https://solo.bodleian.ox.ac.uk/permalink/44OXF_INST/35n82s/alma990174285870107026},
}

@article{arnoldi1951principle,
  title={The principle of minimized iterations in the solution of the matrix eigenvalue problem},
  author={Arnoldi, Walter Edwin},
  journal={Quarterly of applied mathematics},
  volume={9},
  number={1},
  pages={17--29},
  year={1951},
doi={https://doi.org/10.1090/qam/42792}
}

@article{tisseur2001quadratic,
author = {Tisseur, Fran\c{c}oise and Meerbergen, Karl},
title = {The Quadratic Eigenvalue Problem},
journal = {SIAM Review},
volume = {43},
number = {2},
pages = {235-286},
year = {2001},
doi = {10.1137/S0036144500381988},
URL = { https://doi.org/10.1137/S0036144500381988},
eprint = {https://doi.org/10.1137/S0036144500381988}
}

@InProceedings{ghani2020novel,
  author       = {Ghani, Tahira and John Oommen, B},
  booktitle    = {AI 2020: Advances in Artificial Intelligence: 33rd Australasian Joint Conference, AI 2020, Canberra, ACT, Australia, November 29--30, 2020, Proceedings 33},
  title        = {Novel Block Diagonalization for Reducing Features and Computations in Medical Diagnosis},
  year         = {2020},
  organization = {Springer},
  pages        = {42--54},
  doi          = {10.1007/978-3-030-64984-5_4},
  ranking      = {rank5},
}

@article{shukla2022climate,
  title={Climate change 2022: Mitigation of climate change},
  author={Shukla, Priyadarshi R and Skea, Jim and Slade, Raphael and Al Khourdajie, Alaa and van Diemen, Ren{\'e}e and McCollum, David and Pathak, Minal and Some, Shreya and Vyas, Purvi and Fradera, Roger and others},
  journal={Contribution of working group III to the sixth assessment report of the Intergovernmental Panel on Climate Change},
  volume={10},
  pages={9781009157926},
  year={2022},
  publisher={Cambridge University Press Cambridge, UK},
doi={https://doi.org/10.1017/9781009157926}
}

@Article{bavely1979algorithm,
  author  = {Bavely, Connice A. and Stewart, G. W.},
  journal = {SIAM Journal on Numerical Analysis},
  title   = {An Algorithm for Computing Reducing Subspaces by Block Diagonalization},
  year    = {1979},
  number  = {2},
  pages   = {359-367},
  volume  = {16},
  doi     = {10.1137/0716028},
}

@article{malila2024current,
  title     = {Current challenges of alternative proteins as future foods},
  author    = {Malila, Yuwares and Owolabi, Iyiola O and Chotanaphuti, Tanai and Sakdibhornssup, Napat and Elliott, Christopher T and Visessanguan, Wonnop and Karoonuthaisiri, Nitsara and Petchkongkaew, Awanwee},
  journal   = {npj Science of Food},
  volume    = {8},
  number    = {1},
  pages     = {53},
  year      = {2024},
  publisher = {Nature Publishing Group UK London},
}

@article{bravyi2017tapering,
  title={Tapering off qubits to simulate fermionic Hamiltonians},
  author={Bravyi, Sergey and Gambetta, Jay M and Mezzacapo, Antonio and Temme, Kristan},
  journal={arXiv preprint arXiv:1701.08213},
  year={2017},
doi={https://doi.org/10.48550/arXiv.1701.08213}
}

@article{Aitken_1927, 
title={XXV.—On Bernoulli’s Numerical Solution of Algebraic Equations}, volume={46}, 
DOI={10.1017/S0370164600022070}, 
journal={Proceedings of the Royal Society of Edinburgh}, 
author={Aitken, A. C.}, 
year={1927}, 
pages={289–305}
}

@article{Parrish2016,
author = {Parrish, Robert M. and Hohenstein, Edward G. and Martínez, Todd J.},
title = {“Balancing” the Block Davidson–Liu Algorithm},
journal = {Journal of Chemical Theory and Computation},
volume = {12},
number = {7},
pages = {3003-3007},
year = {2016},
doi = {10.1021/acs.jctc.6b00459},
note ={PMID: 27253494},
URL = { https://doi.org/10.1021/acs.jctc.6b00459},
eprint = {   https://doi.org/10.1021/acs.jctc.6b00459}
}

@article{MURRAY1992382,
title = {Improved algorithms for the lowest few eigenvalues and associated eigenvectors of large matrices},
journal = {Journal of Computational Physics},
volume = {103},
number = {2},
pages = {382-389},
year = {1992},
issn = {0021-9991},
doi = {https://doi.org/10.1016/0021-9991(92)90409-R},
url = {https://www.sciencedirect.com/science/article/pii/002199919290409R},
author = {Christopher W Murray and Stephen C Racine and Ernest R Davidson}
}

@article{liu1978simultaneous,
  title={The simultaneous expansion method for the iterative solution of several of the lowest eigenvalues and corresponding eigenvectors of large real-symmetric matrices},
  author={Liu, B},
  journal={Numerical algorithms in chemistry: algebraic methods},
  pages={49--53},
  year={1978},
  publisher={Lawrence Berkeley Laboratory, University of California Berkeley}
}

@article{morgan1986generalizations,
  title={Generalizations of Davidson’s method for computing eigenvalues of sparse symmetric matrices},
  author={Morgan, Ronald B and Scott, David S},
  journal={SIAM Journal on Scientific and Statistical Computing},
  volume={7},
  number={3},
  pages={817--825},
  year={1986},
  publisher={SIAM}
}

@article{schmitz2020quantum,
  title={A quantum solution for efficient use of symmetries in the simulation of many-body systems},
  author={Schmitz, Albert T and Johri, Sonika},
  journal={npj Quantum Information},
  volume={6},
  number={1},
  pages={2},
  year={2020},
  publisher={Nature Publishing Group UK London},
doi={https://doi.org/10.1038/s41534-019-0232-1}
}

@book{gantmakher2000theory,
  title={The theory of matrices},
  author={Gantmakher, Feliks Ruvimovich},
  volume={131},
  year={2000},
  publisher={American Mathematical Soc.}
}

@article{veselie1993jacobi,
  title={A Jacobi eigenreduction algorithm for definite matrix pairs},
  author={Veseli{\'e}, K},
  journal={Numerische Mathematik},
  volume={64},
  number={1},
  pages={241--269},
  year={1993},
  publisher={Springer}
}

@article{BEGOVICKOVAC2024421,
title = {Convergence of the complex block Jacobi methods under the generalized serial pivot strategies},
journal = {Linear Algebra and its Applications},
volume = {699},
pages = {421-458},
year = {2024},
issn = {0024-3795},
doi = {https://doi.org/10.1016/j.laa.2024.07.012},
url = {https://www.sciencedirect.com/science/article/pii/S0024379524003021},
author = {Erna {Begović Kovač} and Vjeran Hari}
}

@article{mcclean2020openfermion,
  title={OpenFermion: the electronic structure package for quantum computers},
  author={McClean, Jarrod R and Rubin, Nicholas C and Sung, Kevin J and Kivlichan, Ian D and Bonet-Monroig, Xavier and Cao, Yudong and Dai, Chengyu and Fried, E Schuyler and Gidney, Craig and Gimby, Brendan and others},
  journal={Quantum Science and Technology},
  volume={5},
  number={3},
  pages={034014},
  year={2020},
  publisher={IOP Publishing}
}

@article{sun2018pyscf,
  title={PySCF: the Python-based simulations of chemistry framework},
  author={Sun, Qiming and Berkelbach, Timothy C and Blunt, Nick S and Booth, George H and Guo, Sheng and Li, Zhendong and Liu, Junzi and McClain, James D and Sayfutyarova, Elvira R and Sharma, Sandeep and others},
  journal={Wiley Interdisciplinary Reviews: Computational Molecular Science},
  volume={8},
  number={1},
  pages={e1340},
  year={2018},
  publisher={Wiley Online Library}
}

@article{saad2023revisiting,
  title={Revisiting the (block) Jacobi subspace rotation method for the symmetric eigenvalue problem},
  author={Saad, Yousef},
  journal={Numerical Algorithms},
  volume={92},
  number={1},
  pages={917--944},
  year={2023},
  publisher={Springer}
}

@article{HARI20141,
title = {Full block J-Jacobi method for Hermitian matrices},
journal = {Linear Algebra and its Applications},
volume = {444},
pages = {1-27},
year = {2014},
issn = {0024-3795},
doi = {https://doi.org/10.1016/j.laa.2013.11.028},
url = {https://www.sciencedirect.com/science/article/pii/S0024379513007623},
author = {Vjeran Hari and Sanja Singer and Saša Singer},
}

@article{stokes2020quantum,
  title={Quantum natural gradient},
  author={Stokes, James and Izaac, Josh and Killoran, Nathan and Carleo, Giuseppe},
  journal={Quantum},
  volume={4},
  pages={269},
  year={2020},
  publisher={Verein zur F{\"o}rderung des Open Access Publizierens in den Quantenwissenschaften},
  doi={https://doi.org/10.22331/q-2020-05-25-269}
}

@article{higham1997stable,
  title={Stable iterations for the matrix square root},
  author={Higham, Nicholas J},
  journal={Numerical Algorithms},
  volume={15},
  number={2},
  pages={227--242},
  year={1997},
  publisher={Springer}
}

@article{higham1986newton,
  title={Newton’s method for the matrix square root},
  author={Higham, Nicholas J},
  journal={Mathematics of computation},
  volume={46},
  number={174},
  pages={537--549},
  year={1986}
}

@inproceedings{wu2021hybrid,
  title={A hybrid approach to joint estimation of MIMO channel and antenna impedance matrices},
  author={Wu, Shaohan},
  booktitle={2021 55th Annual Conference on Information Sciences and Systems (CISS)},
  pages={1--6},
  year={2021},
  organization={IEEE}
}

@article{PEREIRA20011177,
title = {Deflation for block eigenvalues of block partitioned matrices with an application to matrix polynomials of commuting matrices},
journal = {Computers \& Mathematics with Applications},
volume = {42},
number = {8},
pages = {1177-1188},
year = {2001},
issn = {0898-1221},
doi = {https://doi.org/10.1016/S0898-1221(01)00231-0},
url = {https://www.sciencedirect.com/science/article/pii/S0898122101002310},
author = {E. Pereira and J. Vitória},
}

@article{alessandro2023linear,
  title={Linear response equations revisited: A simple and efficient iterative algorithm},
  author={Alessandro, Riccardo and Giann{\`\i}, Ivan and Pes, Federica and Nottoli, Tommaso and Lipparini, Filippo},
  journal={Journal of Chemical Theory and Computation},
  volume={19},
  number={24},
  pages={9025--9031},
  year={2023},
  publisher={ACS Publications},
  doi={http://doi.org/10.1021/acs.jctc.3c00989}
}

@article{Zhong2021Quantum,
  author = {Yuan Zhong and Daniil K. Efetov and Dmitry A. Abanin and Marco Polini and Andrea Young},
  title = {Quantum oscillations in insulating graphene multilayers},
  journal = {npj Quantum Information},
  year = {2021},
  volume = {7},
  pages = {212},
  doi = {10.1038/s41534-021-00416-z},
  url = {https://www.nature.com/articles/s41534-021-00416-z},
}

@article{tong2022decarbonizing,
  title={Decarbonizing natural gas: a review of catalytic decomposition and carbon formation mechanisms},
  author={Tong, Sirui and Miao, Bin and Zhang, Lan and Chan, Siew Hwa},
  journal={Energies},
  volume={15},
  number={7},
  pages={2573},
  year={2022},
  publisher={MDPI},
  doi={https://doi.org/10.3390/en15072573}
}

@article{pelzer2011strong,
  title={Strong correlation in acene sheets from the active-space variational two-electron reduced density matrix method: effects of symmetry and size},
  author={Pelzer, Kenley and Greenman, Loren and Gidofalvi, Gergely and Mazziotti, David A},
  journal={The Journal of Physical Chemistry A},
  volume={115},
  number={22},
  pages={5632--5640},
  year={2011},
  publisher={ACS Publications}
}

@article{cao2023research,
  title={Research progress on graphene production by methane cracking: approach and growth mechanism},
  author={Cao, MJ and Li, SD and Nie, LF and Chen, YF},
  journal={Materials Today Sustainability},
  volume={24},
  pages={100522},
  year={2023},
  publisher={Elsevier},
doi={https://doi.org/10.1016/j.mtsust.2023.100522}
}

@article{patlolla2023review,
  title={A review of methane pyrolysis technologies for hydrogen production},
  author={Patlolla, Shashank Reddy and Katsu, Kyle and Sharafian, Amir and Wei, Kevin and Herrera, Omar E and M{\'e}rida, Walter},
  journal={Renewable and Sustainable Energy Reviews},
  volume={181},
  pages={113323},
  year={2023},
  publisher={Elsevier},
doi={https://doi.org/10.1016/j.rser.2023.113323}
}

@article{macias2023two,
  title={Two global quasi-Newton algorithms for solving matrix polynomial equations},
  author={Mac{\'\i}as, Mauricio and P{\'e}rez, Rosana and Mart{\'\i}nez, H{\'e}ctor Jairo},
  journal={Computational and Applied Mathematics},
  volume={42},
  number={7},
  pages={311},
  year={2023},
  publisher={Springer}
}

@article{leininger2001systematic,
  title={Systematic study of selected diagonalization methods for configuration interaction matrices},
  author={Leininger, Matthew L and Sherrill, C David and Allen, Wesley D and Schaefer III, Henry F},
  journal={Journal of Computational Chemistry},
  volume={22},
  number={13},
  pages={1574--1589},
  year={2001},
  publisher={Wiley Online Library}
}

@article{CAO2023100522,
title = {Research progress on graphene production by methane cracking: approach and growth mechanism},
journal = {Materials Today Sustainability},
volume = {24},
pages = {100522},
year = {2023},
issn = {2589-2347},
doi = {https://doi.org/10.1016/j.mtsust.2023.100522},
url = {https://www.sciencedirect.com/science/article/pii/S2589234723002099},
author = {M.J. Cao and S.D. Li and L.F. Nie and Y.F. Chen},
}

@Article{zhang2022variational,
  author    = {Zhang, Yu and Cincio, Lukasz and Negre, Christian FA and Czarnik, Piotr and Coles, Patrick J and Anisimov, Petr M and Mniszewski, Susan M and Tretiak, Sergei and Dub, Pavel A},
  journal   = {npj Quantum Information},
  title     = {Variational quantum eigensolver with reduced circuit complexity},
  year      = {2022},
  number    = {1},
  pages     = {96},
  volume    = {8},
  doi       = {10.1038/s41534-022-00599-z},
  publisher = {Nature Publishing Group UK London},
  ranking   = {rank5},
}

@Article{STOTSKY2020883,
  author   = {Alexander Stotsky},
  journal  = {IFAC-PapersOnLine},
  title    = {Efficient Iterative Solvers in the Least Squares Method},
  year     = {2020},
  issn     = {2405-8963},
  note     = {21st IFAC World Congress},
  number   = {2},
  pages    = {883-888},
  volume   = {53},
  doi      = {10.1016/j.ifacol.2020.12.847},
  ranking  = {rank5},
  url      = {https://www.sciencedirect.com/science/article/pii/S2405896320311757},
}

@Article{lima2024unitarization,
  author    = {Lima, Dennis and Al-Kuwari, Saif},
  journal   = {Physica Scripta},
  title     = {Unitarization of pseudo-unitary quantum circuits in the S-matrix framework},
  year      = {2024},
  number    = {4},
  pages     = {045202},
  volume    = {99},
  doi       = {10.1088/1402-4896/ad298a},
  publisher = {IOP Publishing},
  ranking   = {rank5},
}
\bibliographystyle{unsrt}

\end{document}